\documentclass[preprint,review,12pt]{elsarticle}
\usepackage{psfrag}
\usepackage{graphicx}
\usepackage{color}
\def\drawline#1#2{\raise 2.5pt\vbox{\hrule width #1pt height #2pt}}

\def\etal{{\it et al. }}

\def\o2{\omega_2}
\def\o1{\omega_1}
\def\o3{\omega_3}
\def\tup{\tilde{u}^\prime}
\def\tupi{\tilde{u}_i^\prime}
\def\tup2{\tilde{u}_2^\prime}

\def\trian{\raise 1.25pt\hbox{$\scriptscriptstyle\triangle$}\nobreak\ }

\def\square{${\vcenter{\hrule height .4pt
        \hbox{\vrule width .4pt height 3pt \kern 3pt
        \vrule width .4pt}
        \hrule height .4pt}}$\nobreak\ }
\def\solidtrian{\raise 1.25pt
\hbox to 3B{
\hfill}\nobreak\ }

\begin{document}

\journal{International Journal of Heat and Fluid Flow}

\begin{frontmatter}



%
\title
{
DNS of laminar-turbulent boundary
layer transition induced by solid
obstacles
}
\author{ P.~ Orlandi S.~ Pirozzoli , \& M. Bernardini }

\address{
Dipartimento di Ingegneria Meccanica e Aerospaziale\\
Universit\`a La Sapienza, Via Eudossiana 16, I-00184, Roma
\\
}


\begin{abstract}

Results of numerical simulations obtained by a staggered
finite difference scheme together with an efficient immersed boundary
method are presented to understand the effects of the 
shape of three-dimensional obstacles on the transition
of a boundary layer from a laminar to a turbulent regime. Fully resolved
Direct Numerical Simulations (DNS), highlight 
that the closer to the obstacle the symmetry 
is disrupted the smaller is the transitional Reynolds number.               
It has been also found that the transition
can not be related to the critical $R_k=U_k k/\nu$ used in
the past.  The simulations highlight the differences 
between wake and inflectional instabilities, proving
that two-dimensional tripping devices are more efficient
in promoting the transition. Simulations at high
Reynolds number demonstrate that the reproduction of
a real experiment with a solid  obstacle at the inlet is an 
efficient tool to generate numerical 
data bases for understanding the physics of
boundary layers. The quality of the numerical method to
fully resolve the small scales, that is one ingredient
for a DNS was shown by a comparison of the exponential range
of  the velocity spectra, in Kolmogorov units, with those for 
isotropic turbulence. The good 
comparison reinforces the idea of local isotropy at the smallest scales.

\end{abstract}

\end{frontmatter}

\section{Introduction}

The laminar-turbulent transition of boundary
layers is of large importance in several practical
applications and in particular  for
the aeronautical industry. Naturally it occurs
when external disturbances affect the near
wall region, generating turbulent spots 
(Emmons \cite{Emm}) occurring randomly
and growing uniformly independently of one another 
as they are swept downstream by the flow.  More
efficiently, the transition can be promoted by inserting solid
obstacles on a smooth wall at a certain distance
from the leading edge. Therefore the shape of the elements may play
an effect on the location at which the flow becomes turbulent. 
The tripping device can be obtained by a series of obstacles
widely separated (thee-dimensional) or by a wire in the whole
span (two-dimensional).  Different scenarios can be achieved,
for  the former configuration the transition can be classified as a wake 
instability, which is associated to the generation of vorticity layers
at the obstacle walls which evolve in the streamwise
direction. For disturbances in the whole span of
the plate an inflectional point  in the velocity profile
inside the recirculating region behind the obstacle appear.
Therefore the small disturbances grow rapidly
and a turbulent flow is generated in the whole
boundary layer. Klebanoff \etal \cite{KCT} 
wrote an exhaustive paper on the transition of boundary layers
induced by solid elements, claiming that two-dimensional obstacles
are  more efficient  to excite three-dimensional disturbances 
than the oscillating ribbon previously used (Klebanoff \etal \cite{KTS}).
The Klebanoff study had  such impact  that 
this type of laminar-turbulent transition was defined as Klebanoff 
transition. 

Klebanoff \etal \cite{KCT} used hemispherical obstacles
with an aspect ratio equal to $1$ and height $k$
smaller than the boundary layer thickness $\delta$. They
compared the values of the transition Reynolds number
($R_k=U_k k/\nu$, where $U_k$ is the velocity of the undisturbed
flow at the top of the element) with those obtained
in other experimental setup.
Large differences were found, for
instance their value equal to $325$ was different than those  in
Tani \etal \cite{TKKI}, varying between $600$ and $1000$.
Dryden \cite{Dr} emphasized that, once the transition occurred, a further
increase of the Reynolds number had the effect of  moving the 
location of the transition upstream.
The reasons of the differences in the critical $R_k$ depend on the nature
of the laminar-turbulent transition.  For instance for isolated
obstacles it  occurs at a certain distance from the obstacle rather smoothly,
and each scholar used different criteria to establish
when the turbulent flow is assessed.
In the discussion of the results Klebanoff \etal \cite{KCT} claimed that
the shape of the obstacle plays a role
and that the data in literature were limited. For us this assertion 
is still true and hence it has been decided to give a contribution
to understand the effects of the shape of the obstacles
by numerical simulations.

The Klebanoff's transition was reconsidered 
experimentally and numerically 
by Bake \etal \cite{BMR}. However, the DNS was 
performed by taking the disturbances from the
experiments and it was observed that
the vortical structures responsible for the transition consisted in 
series of $\Lambda$-shaped vortices.
Ergin \& White \cite{EW}, analysing the experimental results,
related the transitional mechanism to the shape of the
optimal disturbances producing the largest growth
over a specified distance or time. They claimed
that these disturbances are similar to those produced
by three-dimensional obstacles. 
Flow visualizations as   that by Acarlar \& Smith \cite{AS}
have shown that three-dimensional
obstacles in a laminar boundary layer generate 
horseshoe vortices wrapping around the upstream side
of the obstacle with the two legs downstream. 
Flow visualizations are qualitative
and quantitative measurements  are difficult,
in particular to have insights on the effect of
the shape of the obstacle.  Acarlar \& Smith \cite{AS} 
for clear flow visualizations 
used obstacles with $k/\delta=1$ different from those
typical of the tripping devices and different from those 
used by Klebanoff \etal \cite{KCT}. Therefore, they got
different profiles for the streamwise velocity fluctuations. 
In discussing the reason of the differences 
and on the possibility to find a scaling of the transition,
Klebanoff \etal \cite{KCT} were asking for further investigations.
Ergin \& White \cite{EW} designed their experiments
to understand the flow behavior in the Kendall's near and far wake
(Kendall \cite{Ke}). In the near wake region the energy of the disturbance 
decreases;  in the far wake, depending on the Reynolds number,
the growth is similar to the optimal growth predicted by the theory. 
Ergin \& White \cite{EW} used an array of cylindrical
elements separated by a distance such that the elements
can be considered isolated. They did not
change the shape of  the obstacles and they
still used the critical $R_k$ as a parameter to
establish when the flow becomes turbulent.

Our opinion is that $R_k$ can not mark 
the transition from  laminar to  turbulent flow. 
The $R_k$ should be replaced by a quantity related to the
velocity disturbance generated by the solid obstacle.
In wall bounded turbulent flows the normal to the wall
velocity fluctuations ($u_2^\prime$) are those producing the
turbulent Reynolds stress $\langle u_1^\prime u_2^\prime \rangle$
characteristic of the turbulent regime
(the index $i=1$ indicates streamwise, $i=2$ normal and
$i=3$ spanwise directions, angular brackets $\langle  \rangle$
averages in the homogeneous directions
and in time). These fluctuations
produce also $\langle u_2^{\prime 2}\rangle$, the only
normal stress entering in the mean momentum equations, for 
channel and pipes. For boundary layers other 
stresses appear, but the relative terms are one order of magnitude smaller. 
In wall bounded flows the importance of $\langle u_2^{\prime 2}\rangle$ 
was not emphasized in the early experiments, designed to understand the physics 
of the near wall turbulence, for the difficulties to measure $u_2$, 
close to a wall, by hot wires. In numerical simulations
all the quantities are available and in addition the
boundary conditions at the surface can be arbitrarily
varied. For instance, in turbulent channels, Orlandi \etal
\cite{OLTA} demonstrated that, the normal velocity distribution on the
plane of the crests is the driving mechanism for the
modifications of the near wall structures.
The preliminary results suggested that,
a new parametrisation for rough flows could be obtained by
$\tilde{u}_2^\prime|_k$, with
$\tupi=\langle u_i^{\prime 2}  \rangle^{1/2}$
($|_k$ ndicates values at the plane of the crests). 
A continuous transition between smooth ($\tup2 |_k=0$) and rough walls
($\tup2 |_k \ne 0$) was observed. The importance of
this quantity has been further emphasized by Orlandi \cite{O11}
performing  DNS of transitional channels with different type of solid
disturbances in one wall of the channel. It was observed that 
the flow in the channel becomes turbulent when
$\tilde{u}_2^\prime|^+_k> 0.2$ (the superscript $+$ indicates
wall units).  However when this threshold is
achieved, the smallest value for turbulent flows is 
$\tilde{u}_2^\prime|^+_k=0.6$. By increasing the Reynolds number
there is a growth of this stress tending to a value
approximately equal to $1$, which is also the 
value of the maximum $\tilde{u}_2^{\prime^+}$ in smooth pipes or channels.

Bake \etal \cite{BMR} introduced
normal velocity disturbances at the wall derived by laboratory experiments
designed ad hoc. From the previous arguments a similarity between a solid 
obstacle and a distribution of $u_2$ at the wall, for
instance a round jet, does exist.  
Therefore for rough boundary layers $\tilde{u}_2^{\prime}$ could be
a quantity whose value when overcomes a threshold value
detect the laminar-turbulence transition.  Regarding the
transition this quantity can not replace the $R_k$ criterium
because it can not be apriori extimated, but qualitatitevely
it can help to design the optimal shape of the obstacle
producing the highest value at the top, this is out of the
scope of the present paper. 
The strategy to change the shape of the obstacle helps to demonstrate
that $R_k$  can not mark the transition. In fact for
certain geometries the transition may occur and for others no, even
if $R_k$ does not largely varies. 
A further interest is to investigate
whether the numerical simulation of a laminar inlet flow
past a solid obstacle can be an affordable procedure
to get turbulent boundary layers at high Reynolds number. 
To reach these goals it is necessary to understand which tripping device
gives the fastest growth and to have a quantitative
measure of  the transitional distance necessary to get
a fully turbulent regime.
The first DNS of boundary layers by Spalart  \cite{Spa}
used a pseudo-spectral numerical method
together with a fringe method to recycle the output flow
to the inlet. More sophisticated recycling methods were
developed as reported by Pirozzoli \& Bernardini \cite{PBG}
in DNS of compressible turbulent boundary layers. 
Simens \etal \cite{SJHM} asserted that with a distance of
$300$ momentum initial thickness the influence of the artificial 
inlet conditions are eliminated.  Lee \etal \cite{LSK}
used similar inlet conditions in presence of rough walls. 
In these conditions the disturbances, continuously emanating from the interior 
of the rough surfaces, reduce  the distance to achieve a fully turbulent rough 
boundary layer.  Wu \& Moin \cite{WM} simulated the whole transitional
flow by adding to the potential flow patches of isotropic turbulence. 
This disturbance can be considered an attempt  to reproduce 
the conditions in wind tunnels 
similar to those by Emmons \cite{Emm}, where turbulence grows
as random organised turbulent spots. More recently Schlatter \etal \cite{SLBJH} 
used a low-amplitude trip force acting in the wall-normal direction.
In this way they had a rapid laminar-turbulent transition close to the inlet.
The authors did not furnish the velocity distributions so
it is difficult to connect the disturbance to a real tripping device.

With the exception of the DNS of rough boundary layers
(Lee \etal \cite{LSK})  the numerical methods in the 
mentioned simulations did not allow to reproduce the flow interaction
with solid obstacles. The numerical method used in this
paper (Orlandi \& Leonardi \cite{OL}) is based on a second order finite 
difference method and on an immersed boundary technique (IBM) to reproduce
bodies of any shape. The method was tested in previous papers and in 
particular it was shown by Orlandi \etal \cite{OLA} that 
the pressure distribution on the rod-shaped elements
was in a very good agreement with  those measured
by Furuya \etal \cite{FMF}. 
In addition a combined experimental-numerical study
(Burattini \etal \cite{BLOA}) was performed to validate
the DNS, with square bars with $w/k=3$ ($w$ is the spacing between two
elements).  The basic numerical method
requires small modifications to deal with boundary layers.
Radiative outlet boundary conditions allow to the 
vortical structures to exit smoothly from the computational domain. 
The modifications in time and space of the three components of
the vorticity  fields, near the solid element, allows to understand why 
at the same Reynolds number for some geometries are amplified and
for others are not.  A comparison between coarse and refined simulations
(not presented) suggests that the inaccurate representation of the
surface acts as disturbances reducing the critical
Reynolds number, therefore small inperfections in a solid 
obstacle while do not change $R_k$ would enhance the transition,
this is a further reason of the inadequacy of $R_k$ to mark the transition.

\section{
Numerical method }
\label{numerical}

The basic numerical method consists of a second order finite 
difference scheme with staggered velocities, which in the inviscid limit
conserves total energy. The method for the channel is described in
Orlandi \cite{Oa} where the
global conservation properties in presence of non-uniform
grids in the normal directions are presented.  The boundary layers simulations
require inlet, outlet and initial conditions.
The radiative conditions at the outlet 
(Pauley \etal \cite{PMR}) allow the flow vortical structures
to exit from the computational domain without
producing disturbances, which propagating backwards can affect the flow.
In these circumstances, a cos FFT is necessary to have a
direct flow solver for the elliptic equations necessary to have
a solenoidal velocity field. At the inlet a Blasius velocity
profile is imposed by fixing the Reynolds number and the distance 
from the leading edge. The theory of laminar boundary layers
allows to derive at $t=0$ the streamwise and normal velocity profiles
in the entire computational domain. This two-dimensional flow
is replicated in the whole spanwise direction, and 
small random disturbances are added at the inlet to trigger the 
spanwise disturbances, which, interacting with the large scale 
created by the obstacles at the supercritical Reynolds number, 
produce a turbulent flow.

The Navier-Stokes and the continuity equation 
in dimensionless form, are 

\begin{equation}
\frac{\partial
u_{i}}{\partial t} + \frac{\partial u_{i}u_{j}}{\partial x_{j}} =
-\frac{\partial p}{\partial x_{i}}   
+\frac{1}{Re}\frac{\partial^{2} u_{i}}{\partial x_{j}^{2}} \
; \hspace*{1.25cm}
\frac{\partial u_i}{\partial x_{i}}   = 0 \; .
\label{eq1}
\end{equation}

\noindent where $p$ is the dimensionless pressure. 
The Reynolds number is defined as $Re=U_e \delta/\nu$,
being $U_e$ the constant free-stream velocity and
$\delta$ the boundary layer thickness at the inlet section
(usually indicated as $\delta_{99}$) .

The solid element is located at a distance $x_0$ from the inlet,
whose effect on the flow is reproduced by an immersed boundary technique.
The efficiency of the method together with
the MPI (Message Parallel Interface) 
instructions allow to use a sufficient number
of grid points to describe the body surface. The IBM method
differs from that developed by Fadlun \etal \cite{FVOM}, where the
velocities at the nearest points to the surface
were evaluated by interpolations. Indeed Orlandi \& Leonardi \cite{OL},
after the evaluation of the right hand side of the discretized
equations in the regular grid, correct the viscous terms, in the explicit
and implicit step. For the flows here considered
where the obstacles do not move, the metrics coefficients 
at the nearest points are calculated only once  initially.

\section{Results}

\subsection{Low and medium Reynolds number}

A first set of simulations were performed on a coarse grid with $512\times128$ 
points in the streamwise and spanwise directions respectively. In the normal 
direction a non-uniform grid was used with $50$ grid points 
for $ 0< x_2 < 1$ and $50$ for   $1 < x_2 < 10$. 
The three-dimensional solid obstacles have aspect ratio
equal to $1$ and a dimensionless height $k=0.25$.  The computational box, 
for the coarse simulations, has $L_1=25.6$ and $L_3=3.2$. With these parameters
the solid elements are described by $5$ points in $x_1$ and $x_3$
and $25$ points in $x_2$. Such coarse grid was chosen
on the purpose to investigate the dependence of the 
results on the shape of the element.  Three-dimensional 
solid disturbances with cylindrical, cubic and wedge shape were used. 
The experimental results described by Klebanoff \etal \cite{KCT} 
showed that two-dimensional obstacles behave differently than 
three-dimensional obstacles. Therefore an additional set of simulations with
a square rod, located  at the same distance from the 
inlet ($x_0=5$) were performed.

\begin{figure}
\centering
\vskip 0.0cm
\hskip -1.0cm
\psfrag{ylab}{\large $ \omega_3$}
\psfrag{xlab}{\large $ $ }
\includegraphics[height=4.5cm]{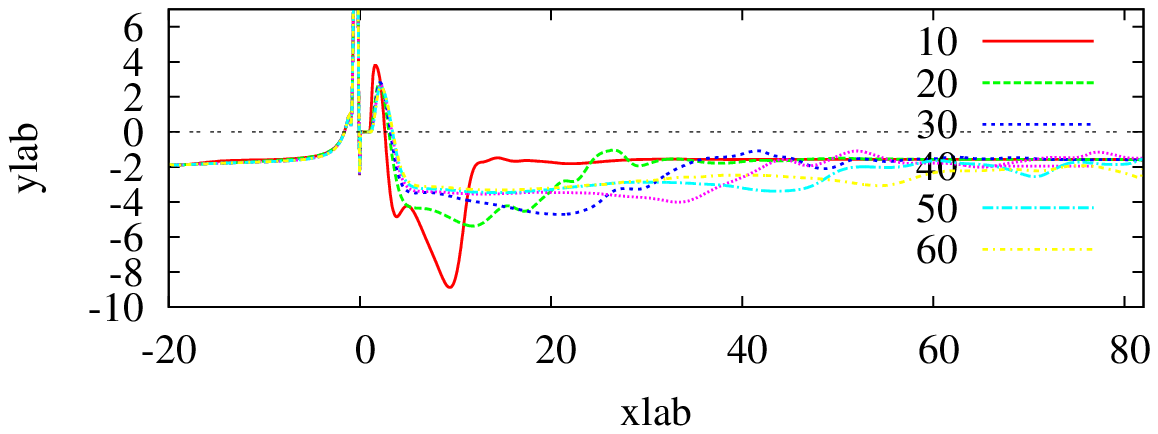}
\vskip -0.3cm \hskip 10cm a) 
\vskip -0.5cm
\hskip -1.5cm
\psfrag{ylab}{\large $ \omega_3$}
\psfrag{xlab}{\large $\eta$ }
\includegraphics[height=4.0cm]{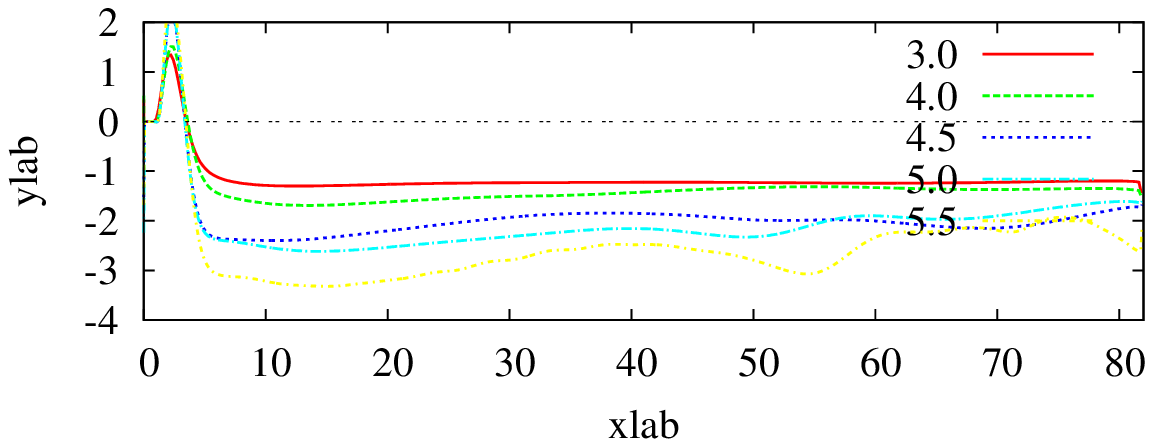}
\hskip -1.0cm
\psfrag{ylab}{\large $ $}
\psfrag{xlab}{\large $\eta$ }
\includegraphics[height=3.5cm]{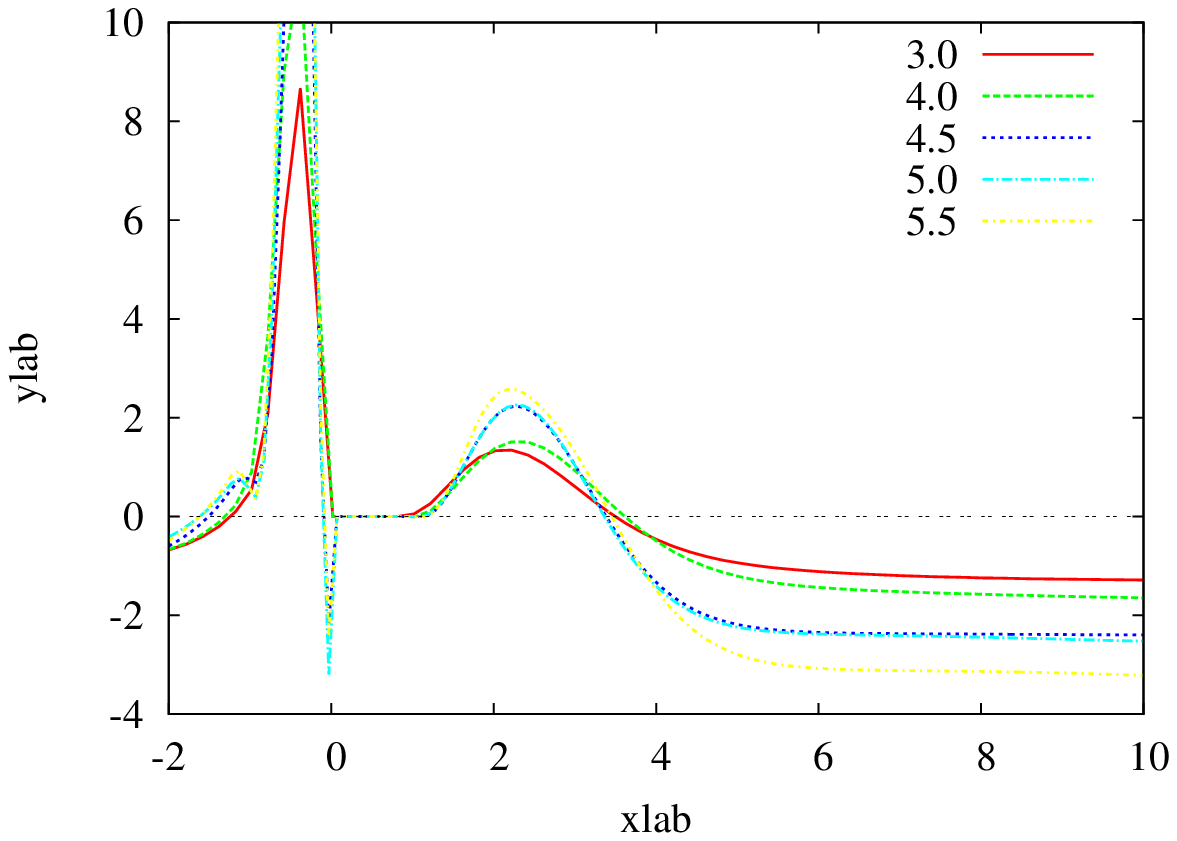}
\vskip -0.3cm \hskip 10cm b) 
\caption{Streamwise profiles of $\omega_3$ 
versus $\eta=(x-x_0)/k$ at 
$y/\delta=0.0041$ at $t=10,20,30,40,50$ and
$60$ for a) the cube at $Re=5500$
$t$ is indicated in the figures,
b) the cube at $t=60$ and differrent $Re 10^3$ as in the inset. 
}
\label{fig1}
\end{figure}

These simulations showed that for the cylinder the  transition
Reynolds number was higher than that for the cubic obstacle.
The latter did not largely differ from that for the wedge.
It was also observed that for square bars the critical Reynolds 
number was much smaller than that for the three-dimensional obstacles.
From these results it was decided to look at the transition
mechanism for the cubic obstacle by varying the $Re$ number
in a well resolved simulations with $1024\times 160 \times 192$
grid points in a domain with the same size above cited. The obstacles
is, then, descibed by $10 \times 15$ grid points in $x_1$ and $x_3$.
The occurrence of transition was detected by monitoring the time evolution
of of $\omega_3$ at $x_2=0.0041$ and $x_3=1.6$.

As it was previously
mentioned, an unsteady simulation evolves towards
a stationary condition, that depending on the value of
$Re,$ remains laminar or becomes turbulent. In both 
cases the transient time necessary to the 
disturbance, produced by the obstacle, to reach the
outlet boundary, depends on $L_1$.  
Figure \ref{fig1}a shows the trends at  $Re=5500$, where
upstream of the obstacle is found a high positive peak, 
due to the formation, of the opposite sign vorticity characterising the 
head of the horseshoe vortex.
Behind and close to the obstacle the other positive peak
is due to the three-dimensional separation bubble, which does not
depend on time, and weakly is affected by $Re$. When the flow reattaches
the wall vorticity varies with $Re$ and with time.
The profiles  at low $Re$ show that 
the negative bump, while is convected downstream, 
decreases its amplitude for viscous  effects, and the
disturbance does not reach the outlet section. Figure \ref{fig1}a
at $Re=5500$ shows that after the short transient, ending at $10<t<20$,
the disturbance travels downstream without an appreciable reduction
for viscous effects.  Figure \ref{fig1}b with the profiles
at different $Re$ show at $Re=4000$ the behavior, similar to that at
$Re=3000$, suggesting a critical $R_{cr}$ close to 
$4500$, which  leads to $R_k=400$
not too different from that observed in the experiments.
The transition starts near the outlet section and
this figure demonstrates that the simulations reproduce the
Dryden \cite{Dr} observations  of the upstream movement
of the transition location with the  increase of $Re$.
The zoom of the region near the obstacle (figure \ref{fig1}b) emphasizes
that, only for $Re>4000$, in front of the obstacle $\omega_3$ is negative
indicating the presence of an intense horseshoe vortex.           

\begin{figure}
\centering
\vskip 0.0cm
\hskip -1.0cm
\includegraphics[width=4.0cm]{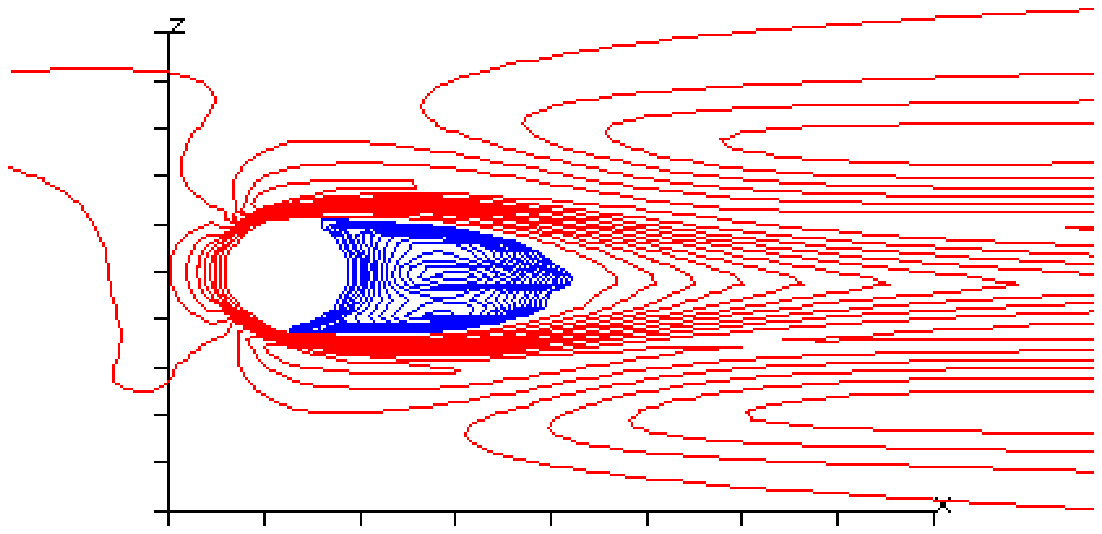}
\includegraphics[width=4.0cm]{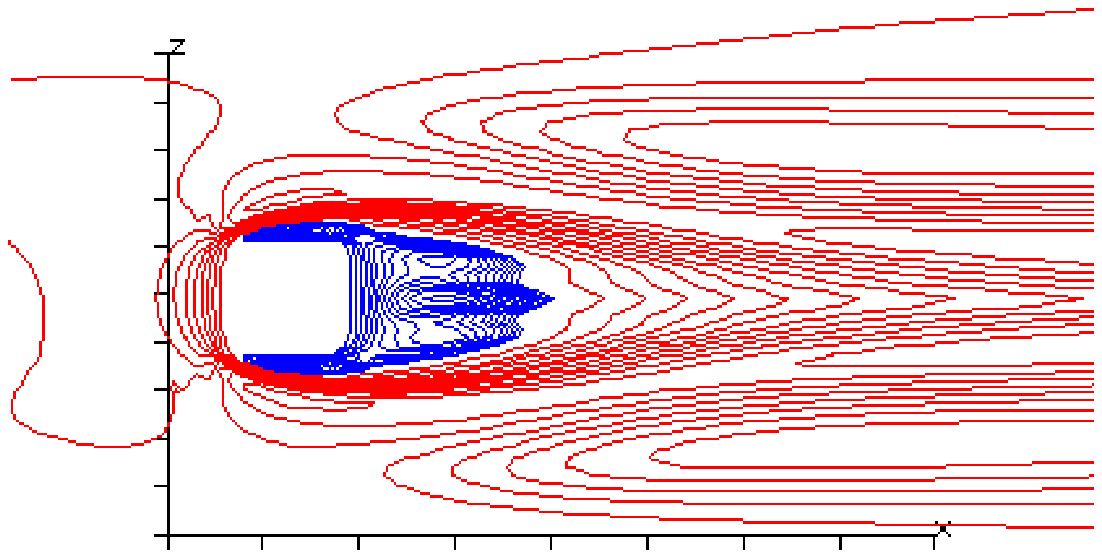}
\includegraphics[width=4.0cm]{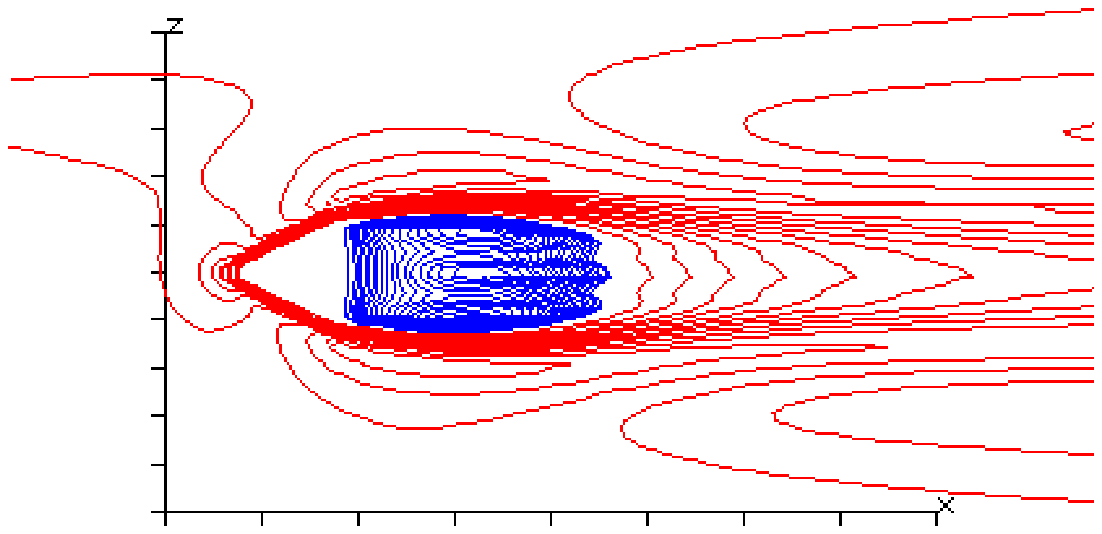}
\vskip -0.3cm \hskip 3cm a) \hskip 3cm b) \hskip 3cm c)
\vskip +0.0cm
\hskip -1.0cm
\includegraphics[width=4.0cm]{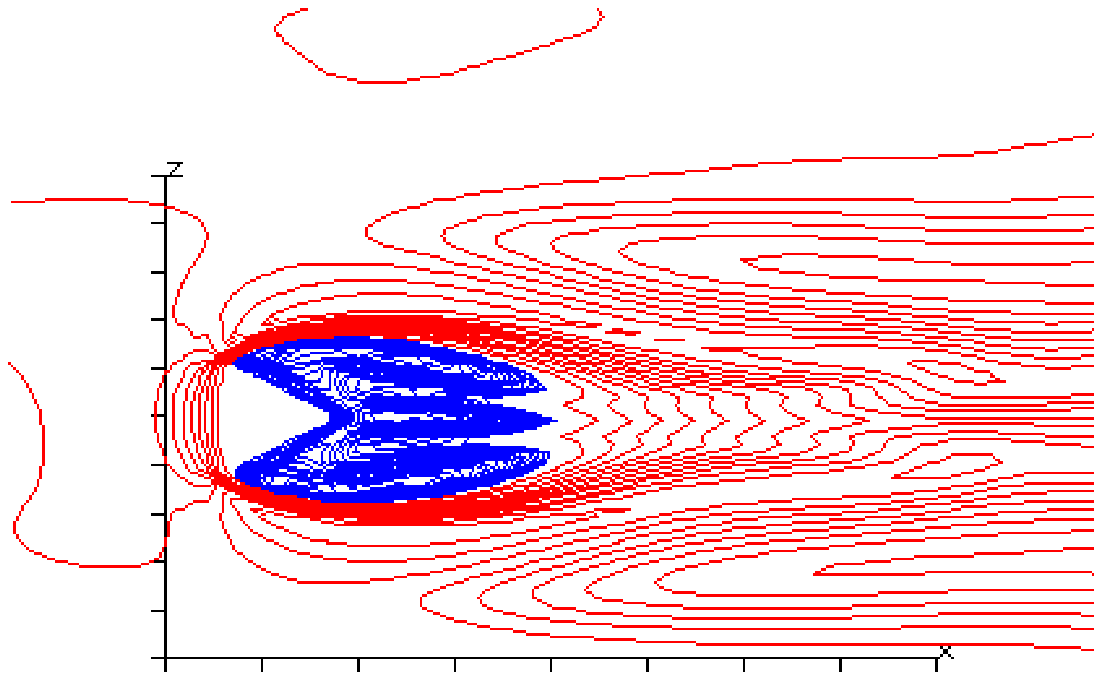}
\includegraphics[width=4.0cm]{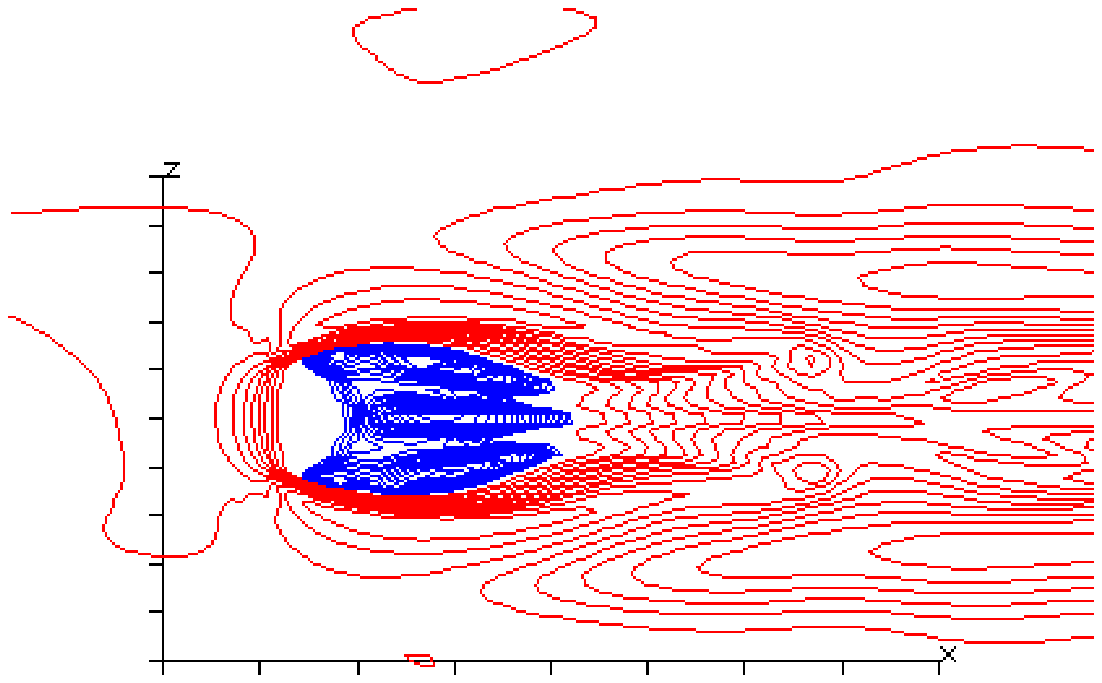}
\includegraphics[width=4.0cm]{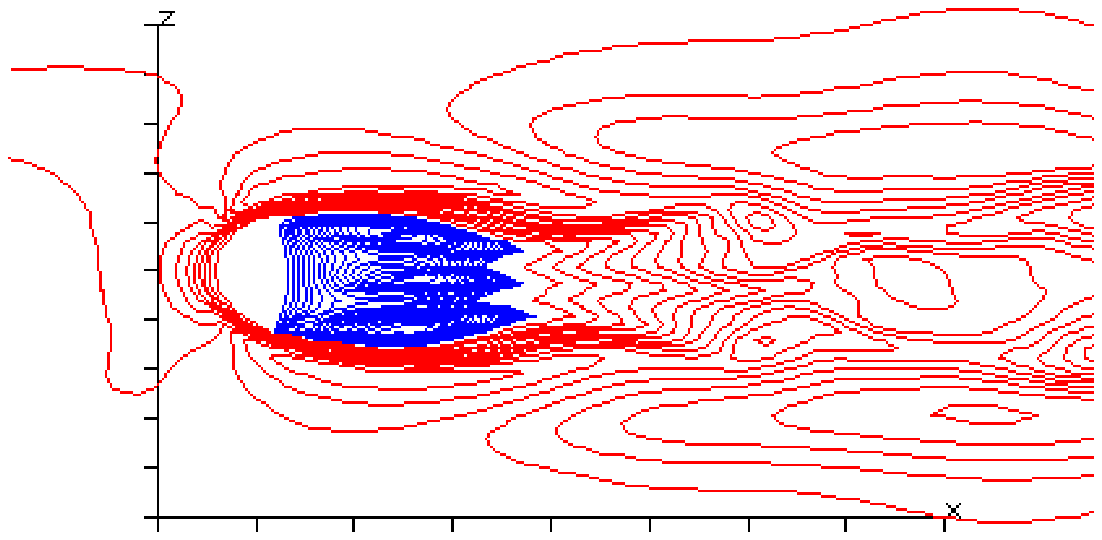}
\vskip -0.3cm \hskip 3cm d) \hskip 3cm e) \hskip 3cm f)
\caption{Streamwise velocity profiles at $y/k=0.6$
for the three-dimensional obstacles simulations
described in a grid with $10$ points in $x_1$
$15$ in $x_3$ and $40$ in $x_2$. The height of
the obstacle is $k=0.25$; red dashed positive, blue solid negative
and $\Delta u_1=0.025$ for the red and $\Delta u_1=0.0025$ 
for the  blue lines.
}
\label{fig3}
\end{figure}

To have a better understanding of the influence of the shape of the 
vortical structures more shapes have been considered at $Re=5500$. 
In figure \ref{fig3} the contour plots of $u_1$, in a small region 
surrounding the body, allow to get an idea of the accurate
reproduction of the flow near the
body. In addition these contours depict the shape of the obstacles considered:
a) cylinder ($CY$), b) cube ($CU$), c) a
wedge pointing upstream ($WU$), d) a wedge pointing downstream ($WD$),
e) half cylinder with straight wall upstream ($SU$) and f) 
with the straight wall downstream ($SD$).
The contours at a distance from the wall 
$x_2=0.15$ highlight the shape of the separation bubble.
The head of the horseshoe vortex does not reach this height  
then does not appear in figure \ref{fig3}. The square bar has been, also, 
considered ($SQ$), but for this two-dimensional obstacle the geometry
is simple and contour plots of $u_1$ are not given. To emphasize the shape 
of the separated region, 
the $\Delta u_1$, for the negative values, has been reduced ten times. 
In figure \ref{fig3} the smooth contours, and, hence, the absence of 
numerical oscillations stress the accuracy of the immersed boundary technique
to reproduce the effects of solid walls.
The different amplitude of the perturbations, in the region
close and downstream of the separation bubble is a first indication that
for the $SU$ and $SD$ obstacles there is the transition.
In addition it can be speculated that for $SD$ the transition starts 
closer to the obstacles.  Similar perturbations are observed for $WD$,
having a smaller amplitude and being more symmetric
the transition moves downstream.
For the the other three geometries no any idea can be drawn
by these images.

\begin{figure}
\centering
\vskip 0.0cm
\hskip -1.0cm
\psfrag{ylab}{\large $ C_f $}
\psfrag{xlab}{\large $\eta$ }
\includegraphics[width=6.5cm]{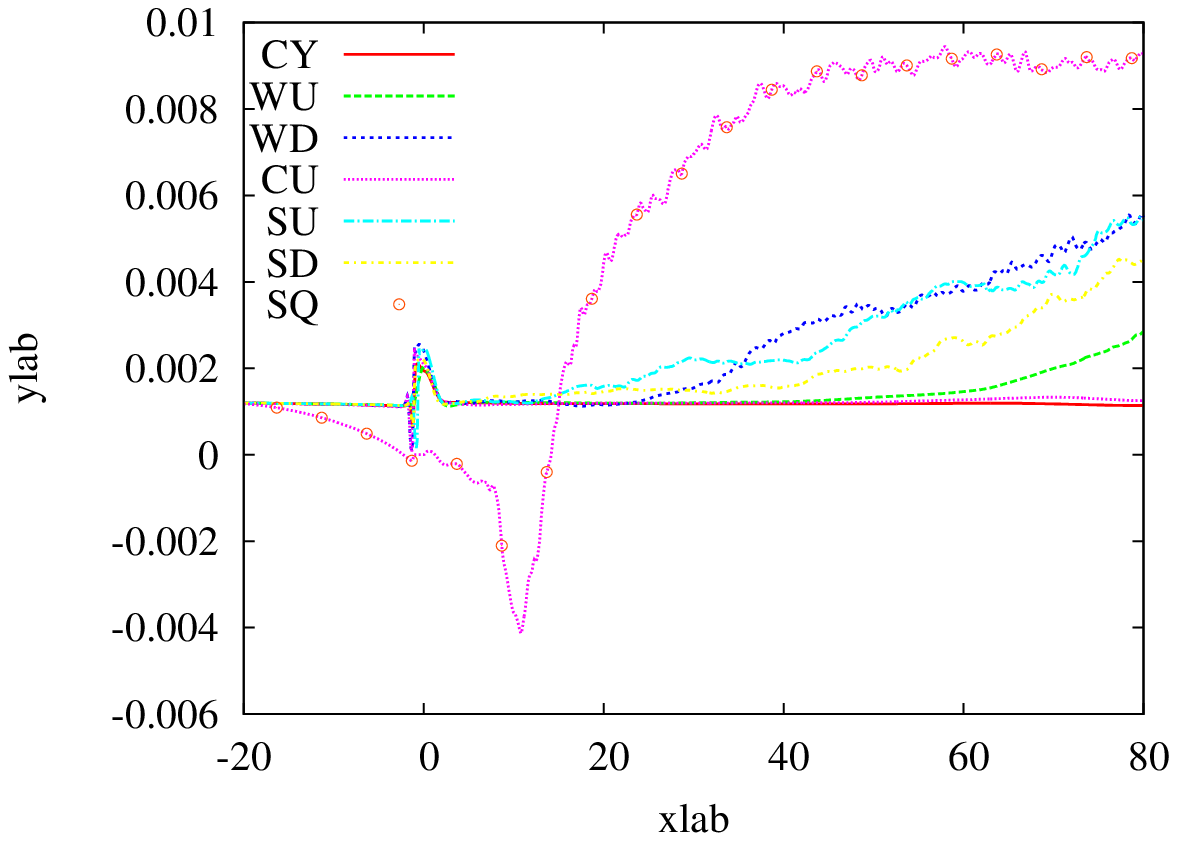}
\psfrag{ylab}{\large $ $}
\psfrag{xlab}{\large $\eta $ }
\includegraphics[width=6.5cm]{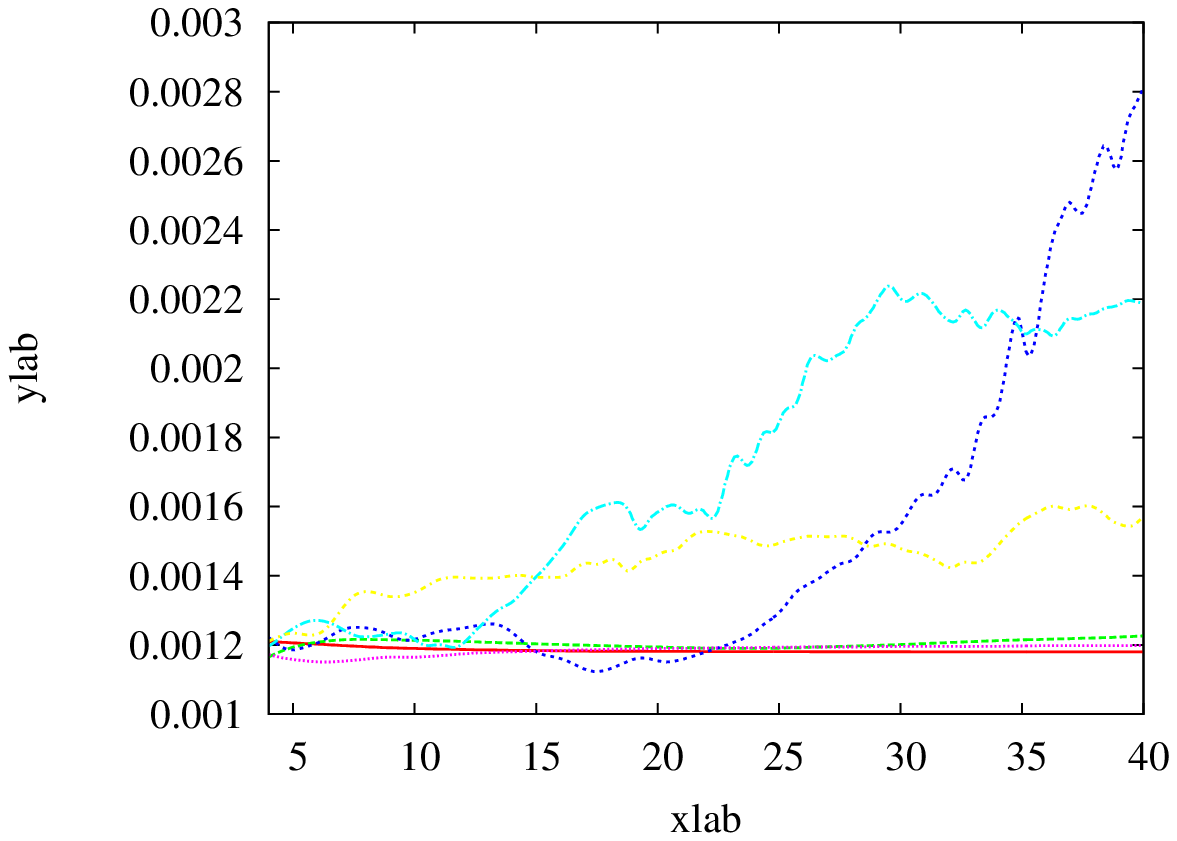}
\vskip -0.3cm \hskip 4cm a) \hskip 5cm b)
\caption{Streamwise profiles of $C_f$ at 
$Re=5500$; a) in the whole length, b) in
a region behind the obstacle to see the location  of
the transition.
}
\label{fig4}
\end{figure}

To have  more quantitative results the spanwise averaged $C_f=2 S_w/Re$
($S_w=\frac {\partial U}{\partial y}|_w$) has been evaluated 
by saving eleven fields from $t=60$ to $t=70$
separated by one time unit. 
It can be argued that a more interesting quantity 
should be the $\omega_3|_w$ at $x_3=1.6$ plotted in figure 1a
with streamwise  oscillations in the turbulent region.
To have smooth profiles
the simulations should evolve for a long time. To detect 
the difference in the growth between the laminar and the turbulent
regime by smoother curves a further averaging in $x_3$ may help. 
For the cylinder in figure \ref{fig4}a
there is no transition.
For the cube the transition  is not close to the obstacle.
To understand the reasons of the transition the surface contours 
of $\omega_3$ and $\omega_2$ may help, as it is discussed later on.
In figure \ref{fig3}b the flow is symmetric with respect
to the centerline, which is an indicator of the absence of
turbulence formation within a reasonable distance from the
obstacle.  A good level of symmetry is 
also obtained for the $WU$ obstacle, where
thin layers of high velocity gradients form (figure \ref{fig3}c). 
These being more unstable, produce the transition at a reasonable distance
from the obstacles, as it results in figure \ref{fig4}a. 
The transition is close to the obstacle for the other three
shapes and to stress the differences, in figure \ref{fig4}b 
the enlargement of the region with the $C_f$ growth confirms the discussion
derived from figure \ref{fig3}. The different
growth is appreciated only at $Re$ close to $R_{cr}$. 
At high Reynolds number these differences
disappear, and, later on, is shown that for the cube, 
the fully turbulent $C_f$ tends to that of a two-dimensional
square bar.  The interesting result of figure 3 is that the
trend of the three-dimensional obstacles is 
different from that of two-dimensional obstacles. 
The transition for the latter flows is  sharp,
indicating a different kind of instability. The 
inflectional point of the $u_1$ profile, in the separation bubble,
causes the growth of the  disturbances in the whole span. 
For the three-dimensional obstacles the disturbances 
are the vortical structures generated by the side walls of the
obstacles, producing other patches of vorticity which
spread laterally to fill-up the entire region. Later on flow visualizations
are used to describe the different physics.

\begin{figure}
\centering
\vskip 0.0cm
\hskip -1.0cm
\psfrag{ylab}{\large $ \tilde{u}_2^\prime|_k $}
\psfrag{xlab}{\large $\eta$ }
\includegraphics[width=6.5cm]{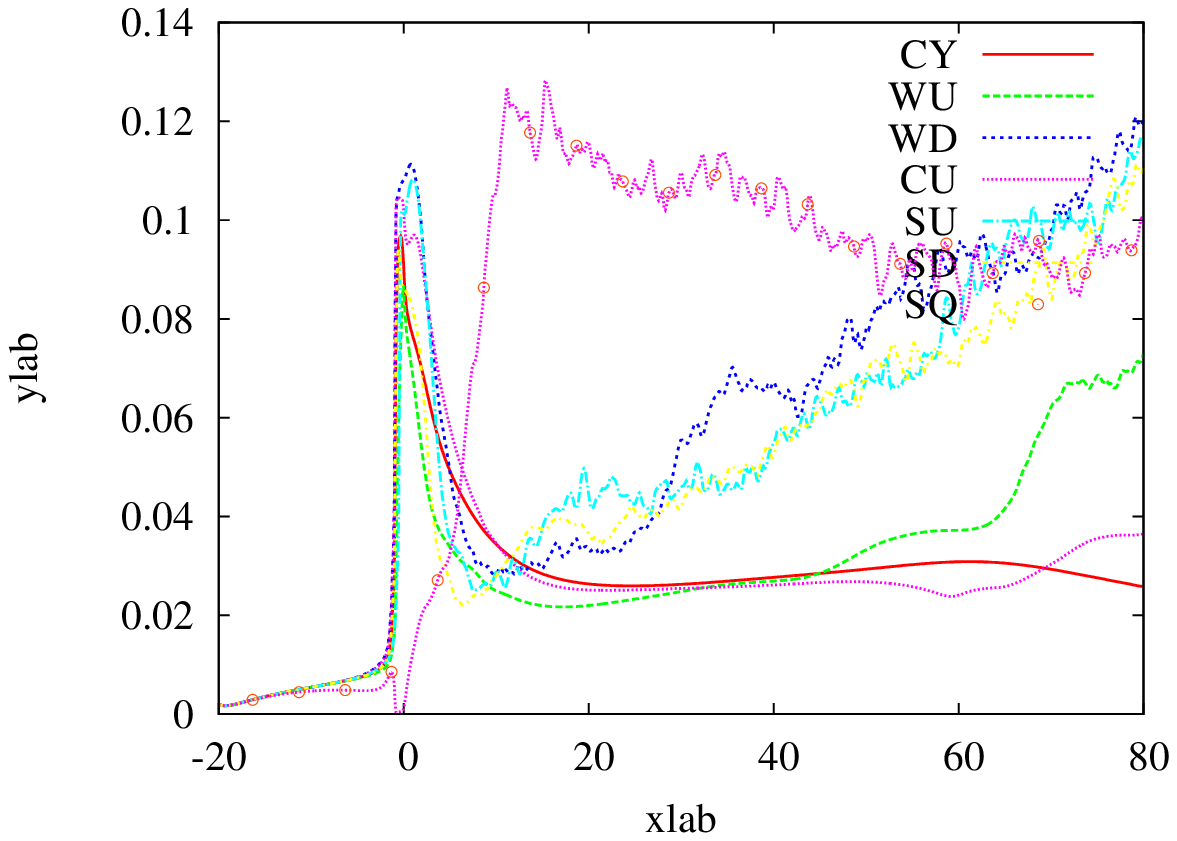}
\psfrag{ylab}{\large $\tilde{u}_2^\prime|_k^+ $}
\psfrag{xlab}{\large $\eta $ }
\includegraphics[width=6.5cm]{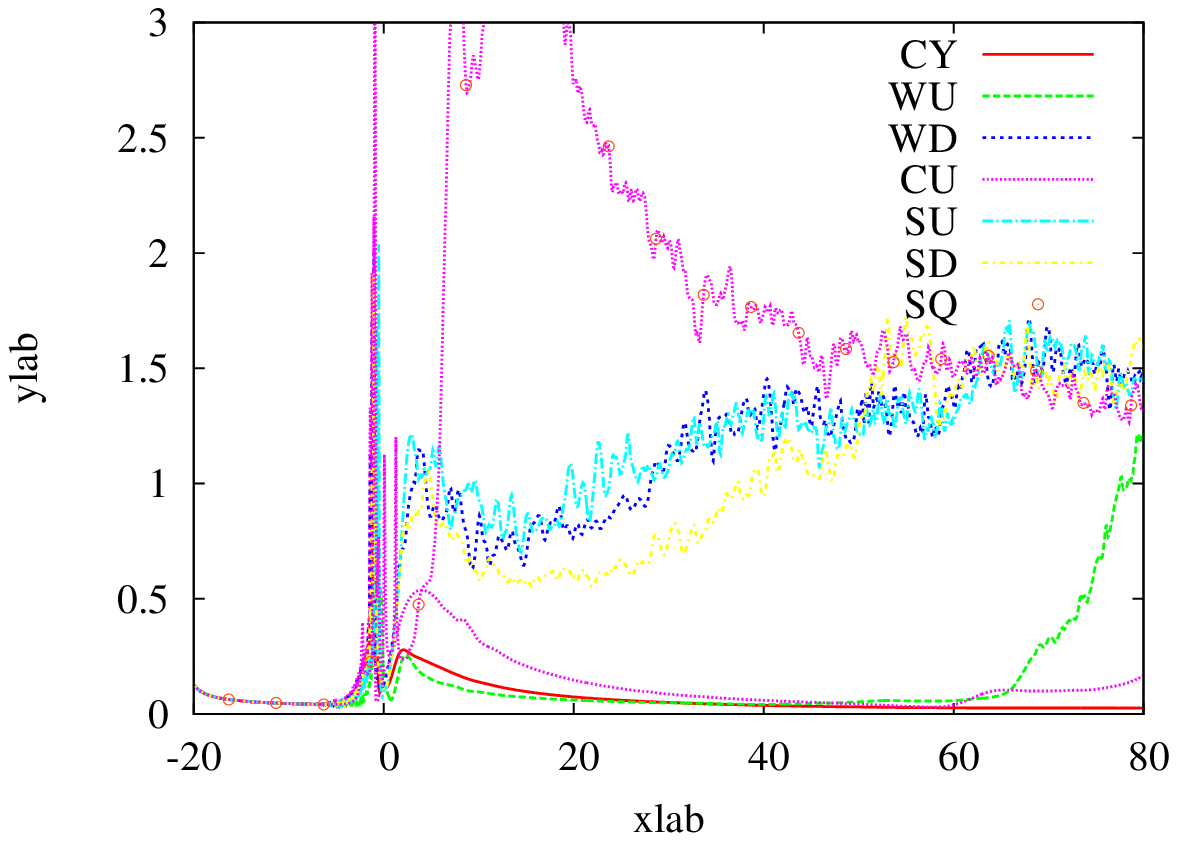}
\vskip -0.3cm \hskip 4cm a) \hskip 5cm b)
\caption{Streamwise profiles of a) $\tilde{u}_2^\prime|_k$ at 
$Re=5500$; b) $\tilde{u}_2^\prime|_k^+$
}
\label{fig5}
\end{figure}

For the three-dimensional obstacles $R_k=U_k k/\nu$ ($U_k$
is the value of $u_1$ at the center of the obstacle and at
$x_2=k$) has been evaluated. Flow visualizations show
small differences of $u_1$ among the different
type of obstacles, with $R_k$ varying between $474$
and $512$, the same range found  by Klebanoff \etal \cite{KCT} 
in several experiments. Since
for some obstacle the transition occurs and for others does
not,  it follows that $R_k$ is not the appropriate
quantity to establish the transition.
A better quantity then should be found.
Orlandi  \cite{O11} in two-dimensional channel demonstrated
that $u_2^\prime$ at the edge of the obstacle produces a 
$\tilde{u}_2^\prime|_k$ varying in time.
After a transient $\tilde{u}_2^\prime|_k$ was rreaching a steady or
unsteady condition depending on the $Re$ number.
In the latter case, charcteristic of a fully turbulent flow,
the averaged $\tilde{u}_2^\prime|^+_k$  in the two homogeneous directions
was greater than a threshold value equal to $0.2$.
For boundary layers $\tilde{u}_2^\prime|_k$ 
is evaluated by averaging in time and in $x_3$. It varies in $x_1$
then it may indicate whether and where the transition occurs. 
Figure \ref{fig5}a shows that in the laminar region, before the obstacle,
with the small disturbances added $\tilde{u}_2^\prime|_k$ is practically 
negligible.  At the obstacle a large jump, related to the thin shear layer 
forming on the top of the obstacle appears.
For viscous effects $\tilde{u}_2^\prime|_k$ decreases downstream and 
for certain obstacles it grows again to lead to the turbulent regime.
For the $WU$ obstacle this happens at $\eta\approx 60$, and, for $CU$, 
it may occur at $\eta >80$.  Figure \ref{fig5}a, however, stresses the 
improbability to have a turbulent regime for $CY$.  For the other 
three-dimensional obstacles a growth similar to that for $WU$
is predicted, which at $\eta\approx 60$ leads to a $\tilde{u}_2^\prime|_k$
greater than that for the two-dimensional obstacle.

Numerical and experimental investigations, by different research groups, 
are currently devoted to understand, in channels, pipes and boundary-layers, 
the reasons of the different trends of the maxima of the
normal stresses with the increase of $Re$. The common view
is that for channels and pipes the dependence is weak
and the value is close to $1$, for boundary layers the
dependence is stronger and the values are slightly higher
( $\approx 1.2$) see Jimenez, \& Hoyas \cite{JH}.  
Figure \ref{fig5}b shows that, at this low $Re$,
despite the different trend, for the three- and the two-dimensional 
obstacles $\tilde{u}_2^\prime|_k^+$ tend towards the value 
typical of boundary layers. 
For the other geometries, generating small disturbances,
$\tilde{u}_2^\prime|_k^+$ decreases for $\eta > 5$, then if  it
increases again and it reaches $\tilde{u}_2^\prime|_k^+ \approx 0.2$ 
it can be asserted that transition to turbulence occurs. 
The  value $0.2$ is close  
to the threshold value found by Orlandi \cite{O11} in transitional channels.
It is important to clarify that for boundary layers over smooth 
walls encountering a single obstacle $\tilde{u}_2^\prime|_k^+$
varies in $x_1$ therefore it can not be considered a threshold
parameter to detect the transition. The result here found
could help in the design of the shape of the obstacle, which
should produce the highest value for $\tilde{u}_2^\prime|_k$.

\begin{figure}
\centering
\vskip 0.0cm
\hskip -1.0cm
\includegraphics[width=6.5cm]{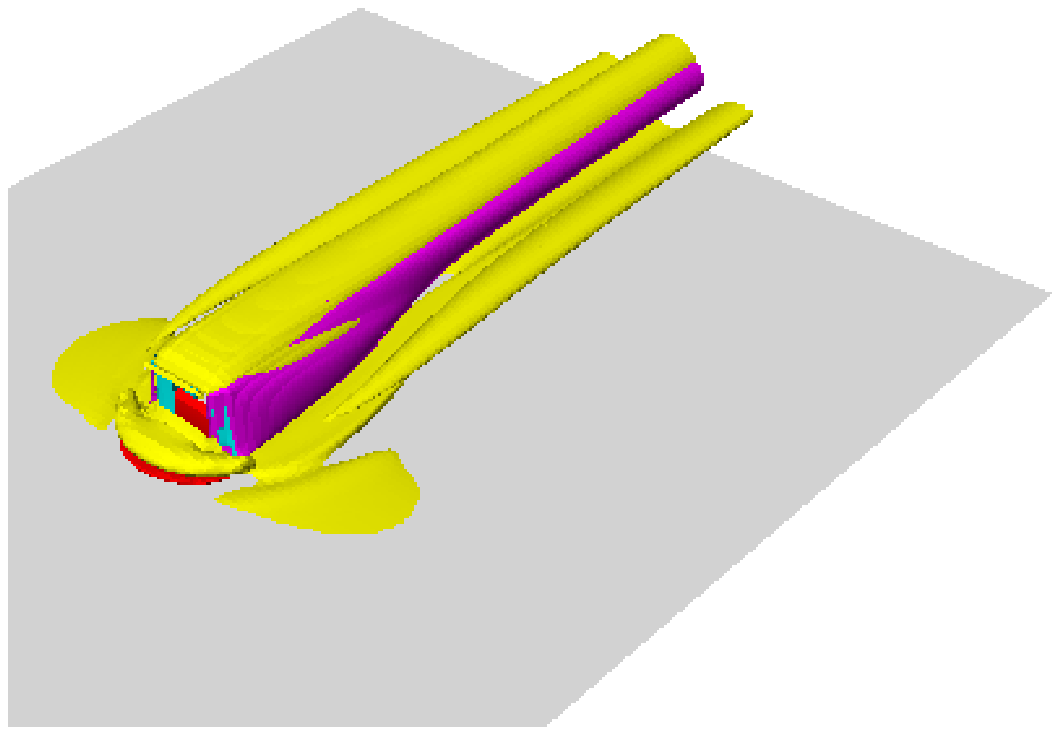}
\hskip 1.0cm
\includegraphics[width=6.5cm]{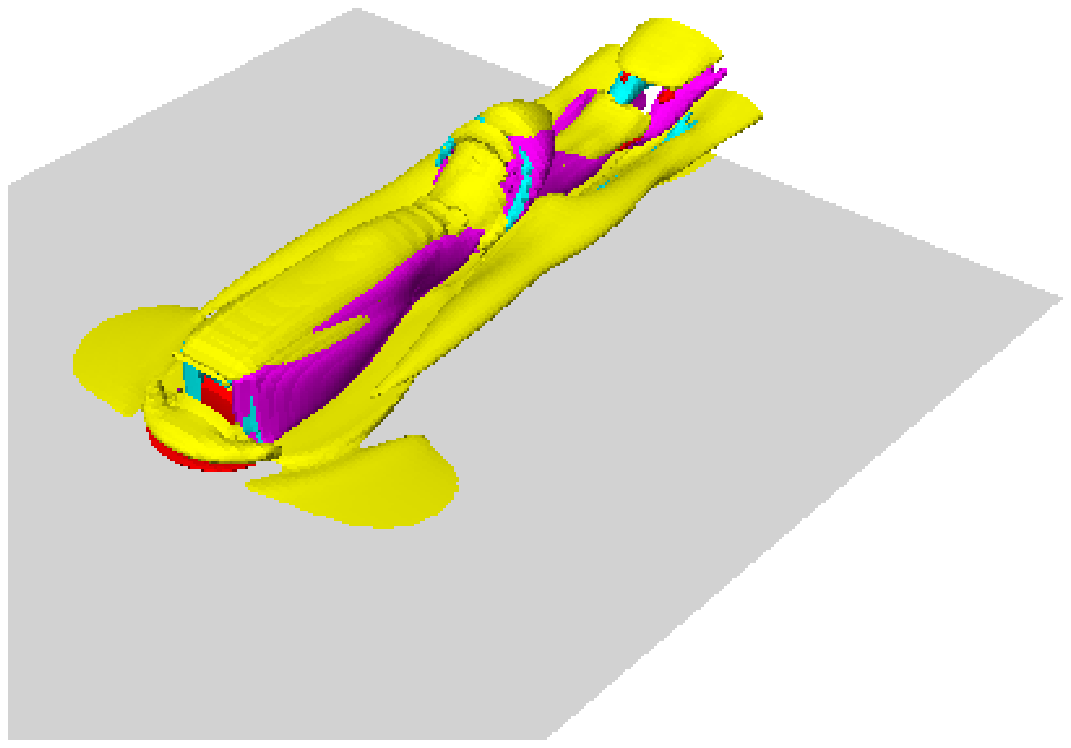}
\vskip -0.3cm \hskip 5cm a) \hskip 5cm b)
\caption{ Surface contours of $\omega_3=-2.5$ yellow 
and $\omega_3=2.5$ red superimposed to $\omega_2=-2$
cyan and $\omega_2=2$ magenta 
a) cube $Re=5500$,
a) $SU$ at $Re=5500$.
}
\label{fig6}
\end{figure}

\begin{figure}
\centering
\vskip 0.0cm
\hskip -1.0cm
\includegraphics[width=6.5cm]{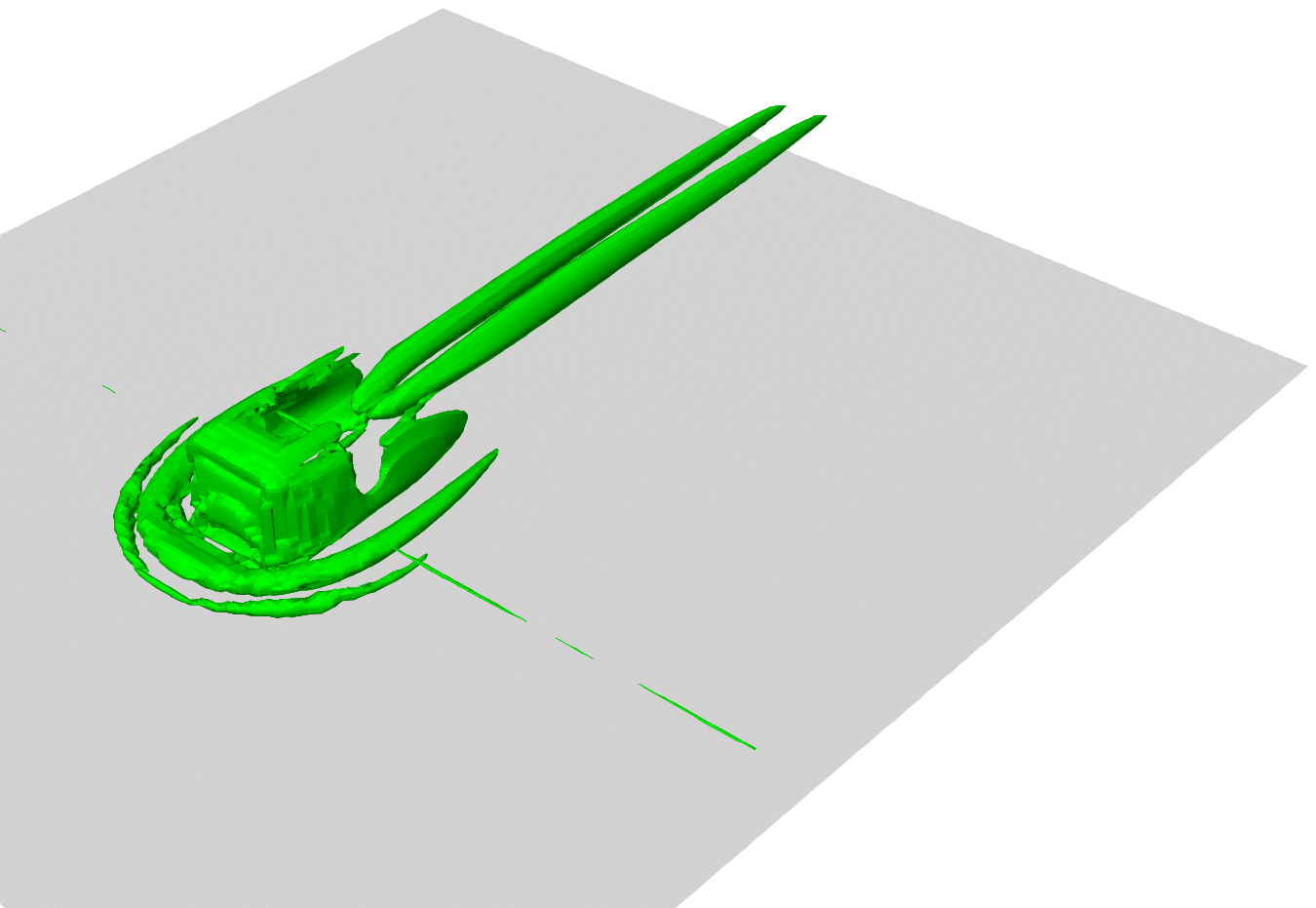}
\hskip 1.0cm
\includegraphics[width=6.5cm]{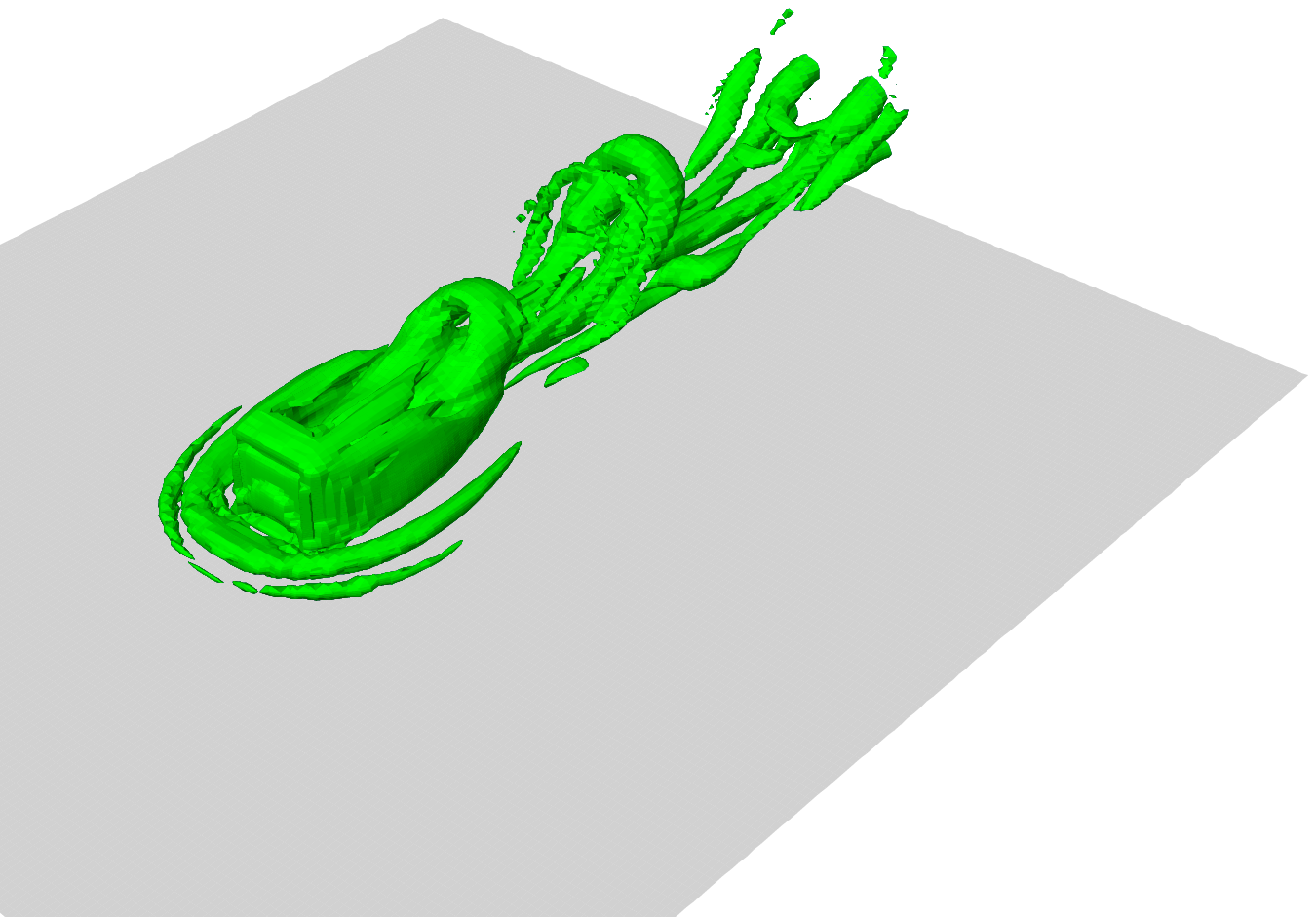}
\vskip -0.3cm \hskip 5cm a) \hskip 5cm b)
\caption{ Surface contours 
of surface contour of $\lambda_2=1$
a) cube $Re=5500$,
a) $SU$ at $Re=5500$.
}
\label{fig7}
\end{figure}

\begin{figure}
\centering
\vskip 0.0cm
\hskip -1.0cm
\includegraphics[width=6.5cm]{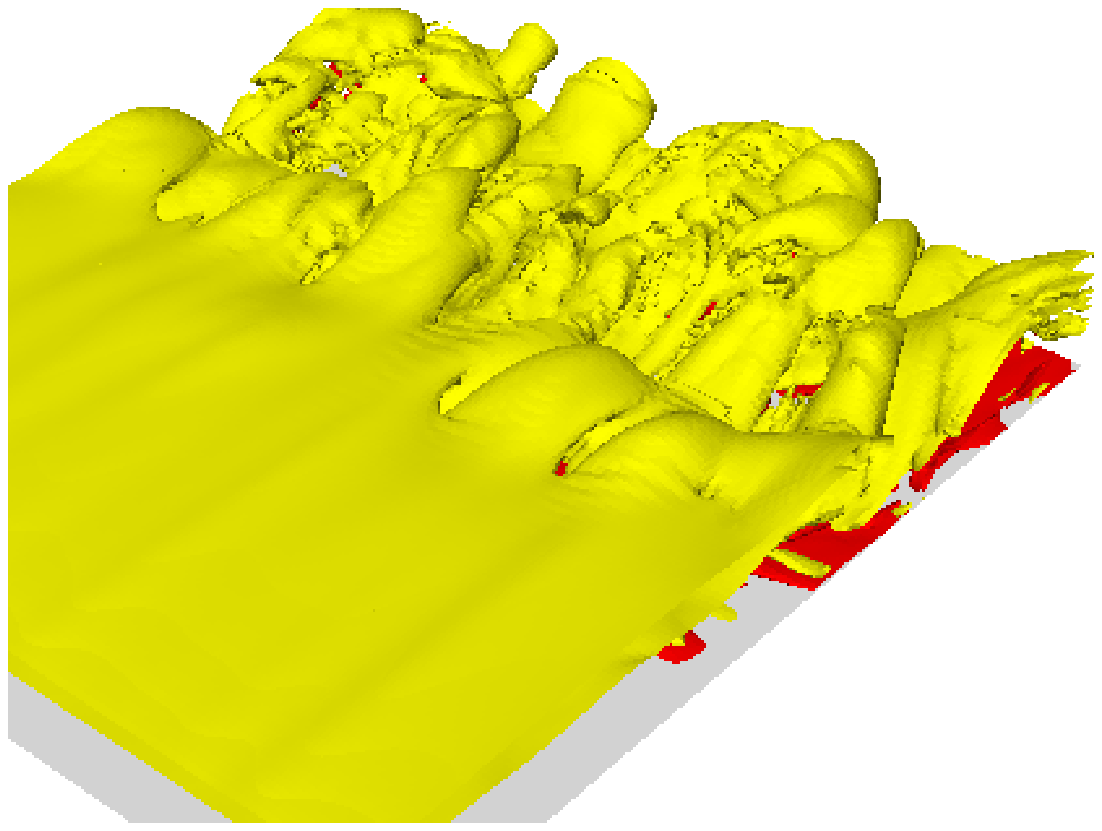}
\hskip 1.0cm
\includegraphics[width=6.5cm]{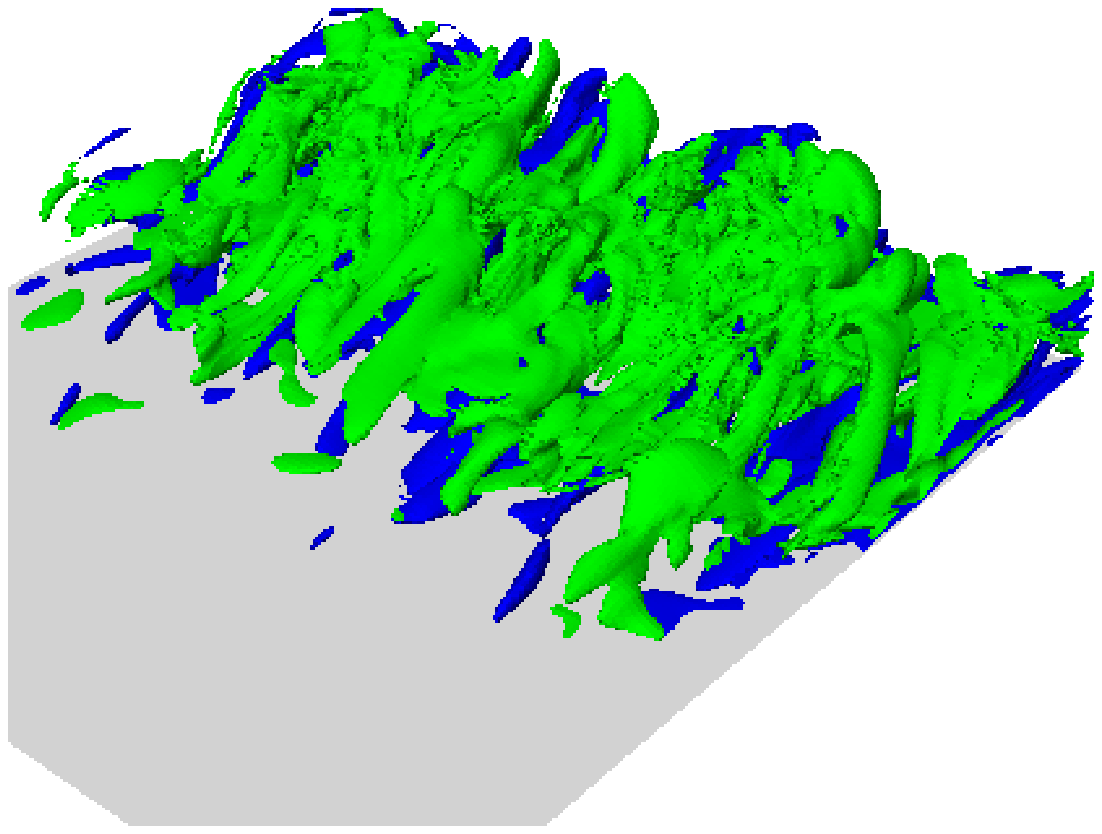}
\vskip -0.3cm \hskip 5cm a) \hskip 5cm b)
\vskip 0.5cm
\hskip -1.0cm
\includegraphics[width=6.5cm]{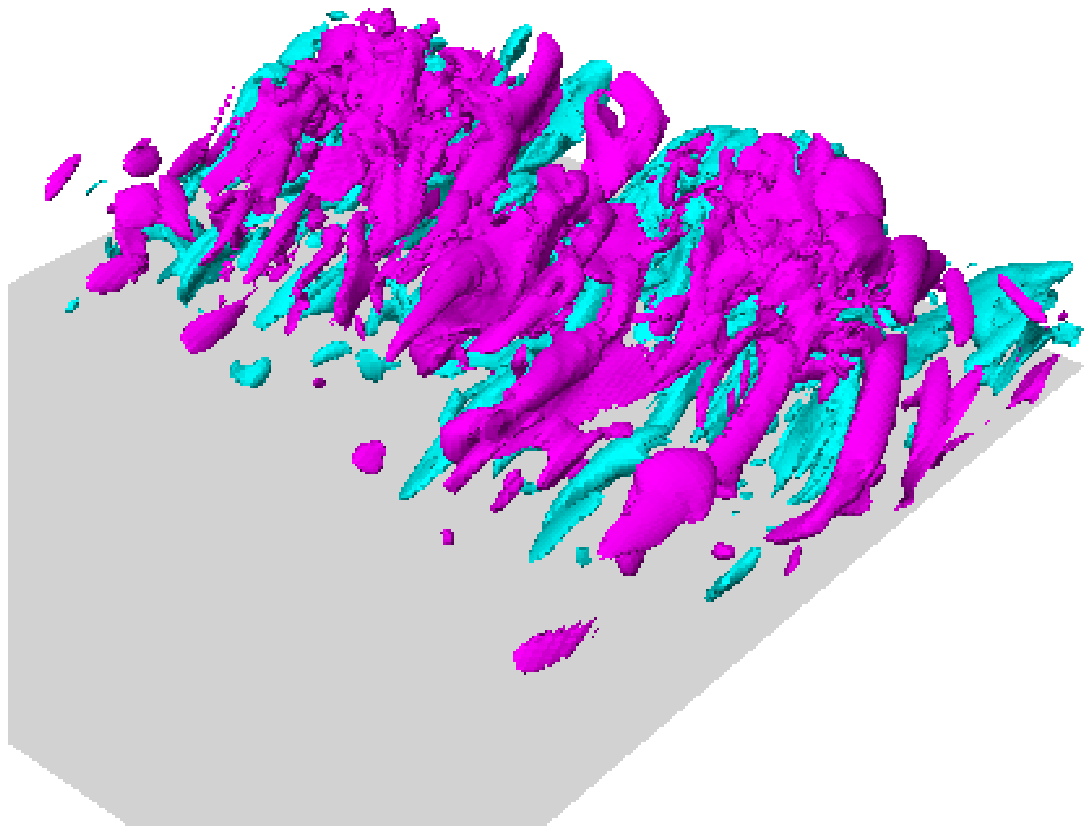}
\hskip 1.0cm
\includegraphics[width=6.5cm]{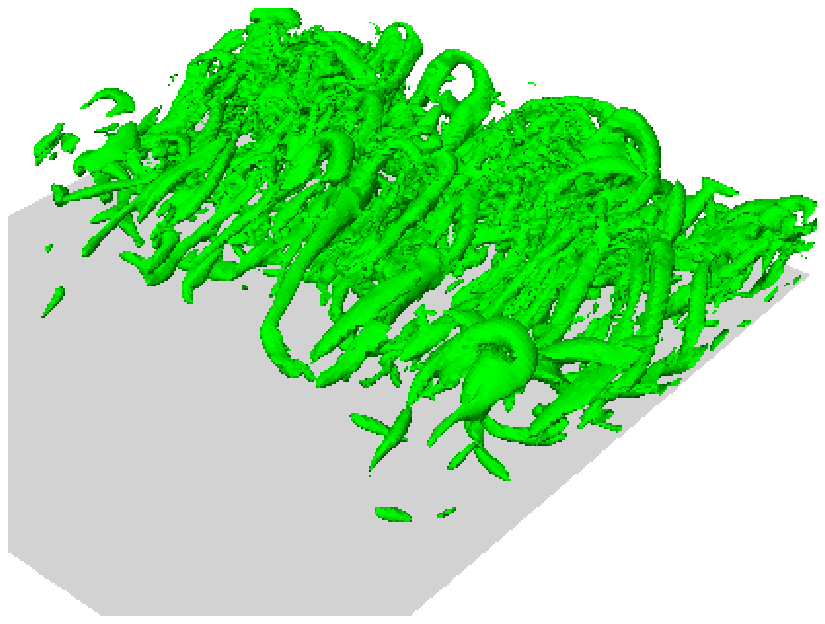}
\vskip -0.3cm \hskip 5cm c) \hskip 5cm d)
\caption{ Surface contours for the case 
$SQ$ at $Re=5500$:
a) $\omega_3=-2.5$ yellow and $\omega_3=2.5$, 
b)   $\omega_1=-2$ blue and $\omega_1=2$ green 
c)   $\omega_2=-2$ cyan and $\omega_2=2$ magenta 
d)  $\lambda_2=2$.
}
\label{fig8}
\end{figure}

Flow visualizations of the vorticity field surrounding
the obstacle may explain  the different
behaviors discussed for the three-dimensional obstacles. The
$CU$ and the $SU$ cases are analysed, the first
does not lead to transition near the obstacle, the latter does.
The visualization in figure \ref{fig6}a and figure \ref{fig6}b emphasizes 
the important role of $\omega_2$,
related to  the spanwise gradients of $u_1$
in presence of three-dimensional obstacles.  
Figure \ref{fig6}a
does not produce a sinuous instability on $\omega_2$, the thin $\omega_2$
layers roll-up remaining stable for a long distance. 
The head of the horses shoe vortex (red structure) at the bottom of the  
front face does not play a role.  This figure
leads to conclude that in order to promote the
transition the shape of the obstacle should create vortical structures with
a strong tendency  to become unstable. 
Figure \ref{fig3} showed that for the half cylinders ($SU$) the thin 
regions with a clustering of $u_1$ isocontours are  curved and consequently
unstable.  The comparison between figure \ref{fig6}b and
figure \ref{fig6}a shows equal vorticity layers in front
of the obstacle.  The circular shape of the side wall promotes the 
instability and the breakdown of the $\omega_2$ and $\omega_3$ layers.
Flow structures are, usually, visualised by the $\lambda_2$
surface contours (the imaginary part
of the complex conjugate eigenvalue of the velocity
gradient tensor), in figure \ref{fig7}a and figure \ref{fig7}b this 
quantity highlights the horseshoe vortex and a series
of hairpins downstream the obstacle for the $SU$, while 
for the $CU$ there is a similar horseshoe vortex and
two circular vortices behind the obstacle. These structures
have been visualised and studied by Acarlar \& Smith \cite{AS}.
More recently a forest of hairpins were visualised
in the DNS of transitional boundary layers (Wu \& Moin \cite{WM}).
The  $\lambda_2$  detects the tubular structures which 
are the structures less important to drive the instability
and the energy cascade characteristic of turbulent flows.
Figure \ref{fig6} together with figure \ref{fig7} 
indicate that the rollup of the $\omega_2$ 
layers contributes to the legs, the rollup of the $\omega_3$ sheets 
to the head of the hairpins.

These visualizations are useful to highlight the differences among
the three- and the two-dimensional obstacles. A
square bar creates a thin vorticity layer at its
top and a recirculating bubble below it. 
Figure \ref{fig8}a shows the formation of the thin $\omega_3$ 
layer bending and breaking at a certain $x_1$.
For the effect of the other two vorticity components
the $\omega_3$ layers form structures of circular shape.
For two-dimensional obstacles $\omega_1$  and $\omega_2$ are
similar (figure \ref{fig8}b and
figure \ref{fig8}c). These components are generated as 
fast growing instabilities due to the inflectional point in the
profiles of $u_1(x_2)$. The forest of hairpins 
depicted in figure \ref{fig8}d is similar to that
in Wu \& Moin \cite{WM}. From these visualizations
the reasons of the  sharp growth of
the $C_f$ in figure 4 are understood. The forest of hairpins
is observed also by three-dimensional obstacles,
but at large distances from the solid element.   
To investigate the analogies and the differences
between two- and three-dimensional obstacles simulations 
at a higher $Re$ are presented.

\subsection{High Reynolds number}

Having demonstrated that the shape of the obstacle 
affects the transition, it is worth  to investigate
whether the simulations with the tripping devices can produce 
turbulent boundary layers at a reasonable high Reynolds number. 
In the last decade the possibility to use clusters with a large number 
of processors allowed to get DNS competing with laboratory experiments. 
Hoyas  \& Jimenez \cite{HJ}, for plane channels, had results  
up to $R_\tau=2000$.  This $Re$ is smaller than those in laboratory
experiments, nevertheless for the difficulty  to have accurate measurements at 
$R_\tau=2000$ the results were useful to the fluid dynamics community.
In  particular for the scholars interested to the $\tilde{u}_2^\prime$,
to the vorticity rms, and to high order statistics. 
Regards boundary layers, the largest $R_\tau$ was 
simulated by Schlatter \etal \cite{SLBJH}
at $R_\theta=4300$ ($R_\theta$ is the Reynolds number
based on the momentum thickness) and results at higher $R_\theta$,
in progress, were presented at conferences. For boundary
layers it is easier to reach high $R_\theta$ in laboratories, as
reported in the review by Smits \etal \cite{SMM}.  However, the numerical 
simulations helped to understand the reasons of certain  erroneous 
conclusions due to the limitations of the size of the probes. 
The present simulations reached   $R_\theta \approx 1500$, and,
to our knowledge are the first one to reproduce the
turbulent boundary layers generated by tripping devices.

Simulations for a square bar and for a cube were performed 
at $Re=7500$ in a domain with $L_1=51.2$ and $N_1=1024$;
$L_3$ was the same as in the previous section but the number of points
in $x_3$ were doubled. In the normal direction the resolution was the same,
in fact with $160$ points, close to  the outlet section, the first grid point
is at $y^+=0.845$.  The comparison between the square bar and the cube 
showed  that, even for the square bar, only for certain statistics, a fully 
turbulent condition was achieved. For the square bar a further
simulation more resolved in $x_1$  and in a longer domain ($L_1=68.2$) 
($N_1=2048$) was performed. In the normal directions the better
resolution was obtained with $L_2=9$ and $192$ points.
In wall units the resolution in the streamwise and spanwise 
directions, in the fully turbulent region, is respectively
equal to $10.5$ and $2.7$.

\begin{figure}
\centering
\vskip 0.0cm
\hskip -1.0cm
\psfrag{ylab}{\large $ H_{12} $}
\psfrag{xlab}{\large $R_\theta$ }
\includegraphics[width=6.5cm]{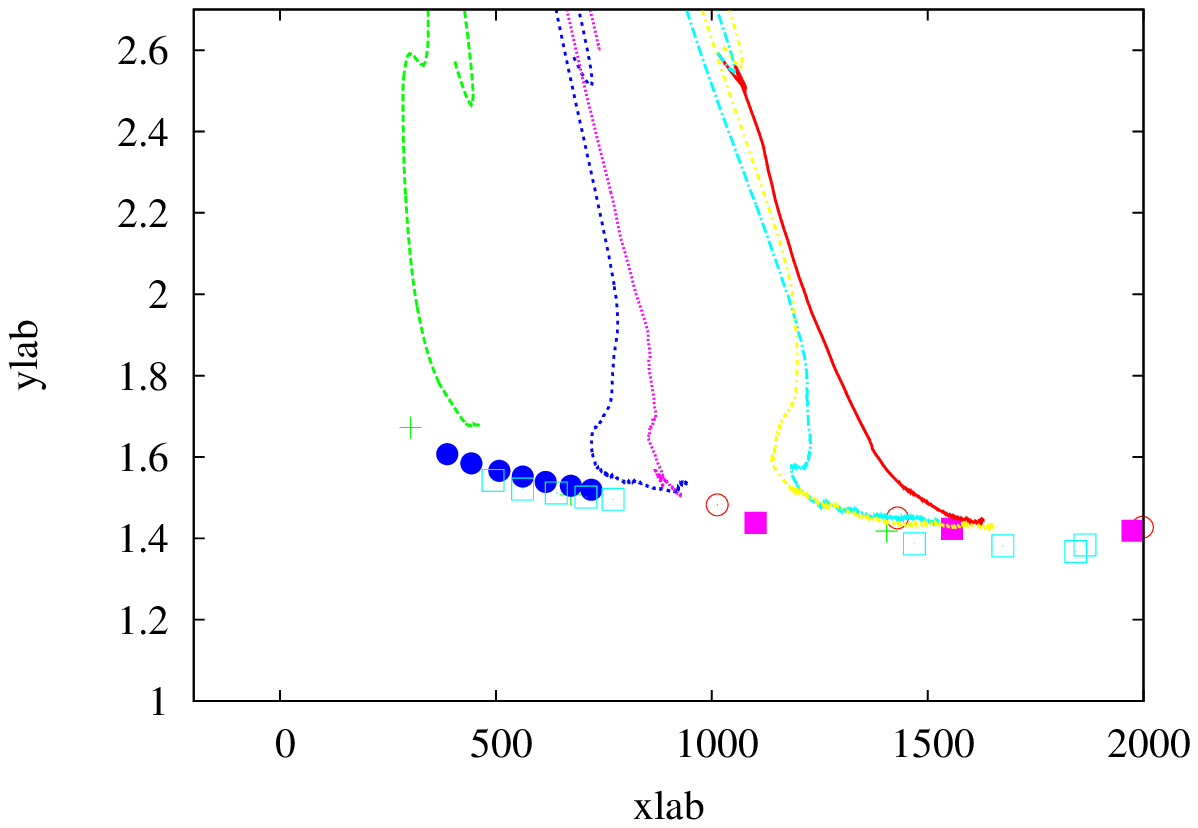}
\hskip 0.0cm
\psfrag{ylab}{\large $ C_f $ }
\psfrag{xlab}{\large $R_\theta$ }
\includegraphics[width=6.5cm]{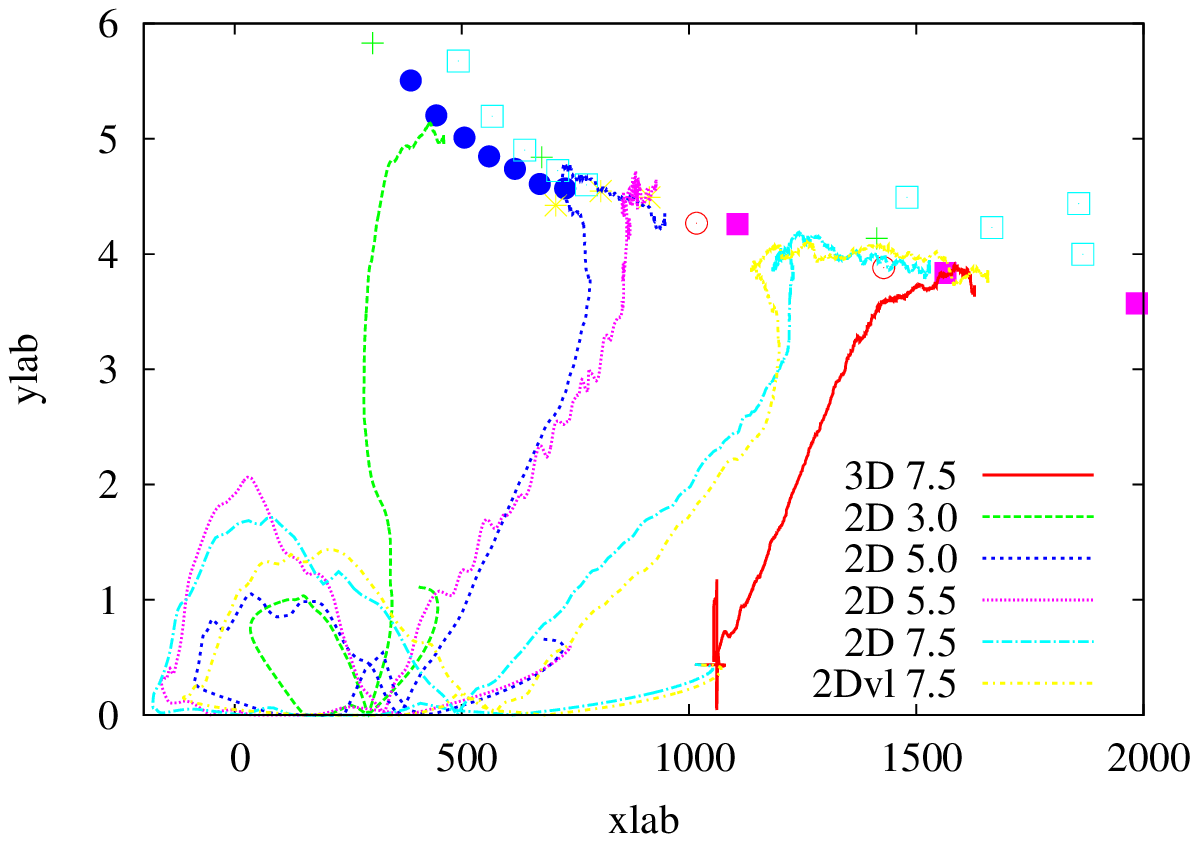}
\vskip -0.3cm \hskip 5cm a) \hskip 5cm b)
\caption{ a) Shape factor $ H_{12} $, b)
friction factor $C_f$ as function of $R_\theta$, the symbols
are the data in Schlatter \&  Orlu \cite{SO} and the
lines the present simulations as indicated in
the insert.
}
\label{fig9}
\end{figure}

Schlatter \&  Orlu \cite{SO} compared the variations of the shape 
factor $H_{12}$ and $C_f$ with $R_\theta$ with those by other simulations,
showing discrepancies but generally the same trend. They reported the 
values of the fully turbulent regime. The present simulations,
including the transition, highlight the trend towards the fully 
turbulent regime.  Figure \ref{fig9}a  reproduces the decrease of $H_{12}$ 
from the laminar to the turbulent value, which reaches the values  of 
Schlatter \&  Orlu \cite{SO} at different $R_\theta$.  For the 
three-dimensional obstacles the values with the cube at 
$Re=7500$ show that the trend for the cube differs from that, 
at the same $Re$, of the square bar.
For the latter geometry small differences in the transitional regime are
observed, but all the simulations in the fully turbulent regime give
$H_{12}$  in good agreement with that of Schlatter \&  Orlu \cite{SO}.
Figure \ref{fig9}b shows a similar agreement for 
the $C_f$. Here the negative values for $R_\theta$ account for the
separation bubble.  For the square bars at $Re=7500$ 
small differences in the $C_f$  are encountered in the transitional regime,
but independently from the resolution ($vf$ in the insight of figure \ref{fig9}b
indicates the more resolved) in the fully turbulent regime the 
agreement with Schlatter \&  Orlu \cite{SO} data extends for a wide range 
of $R_\theta$ values.

\begin{figure}
\centering
\vskip 0.0cm
\hskip -1.5cm
\includegraphics[width=8.0cm]{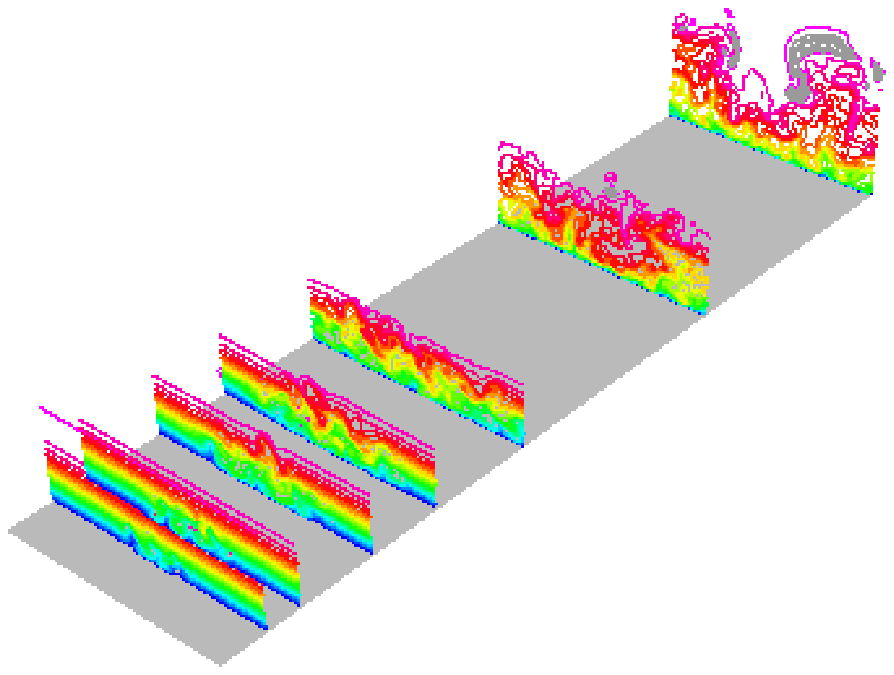}
\hskip -1.2cm
\includegraphics[width=8.0cm]{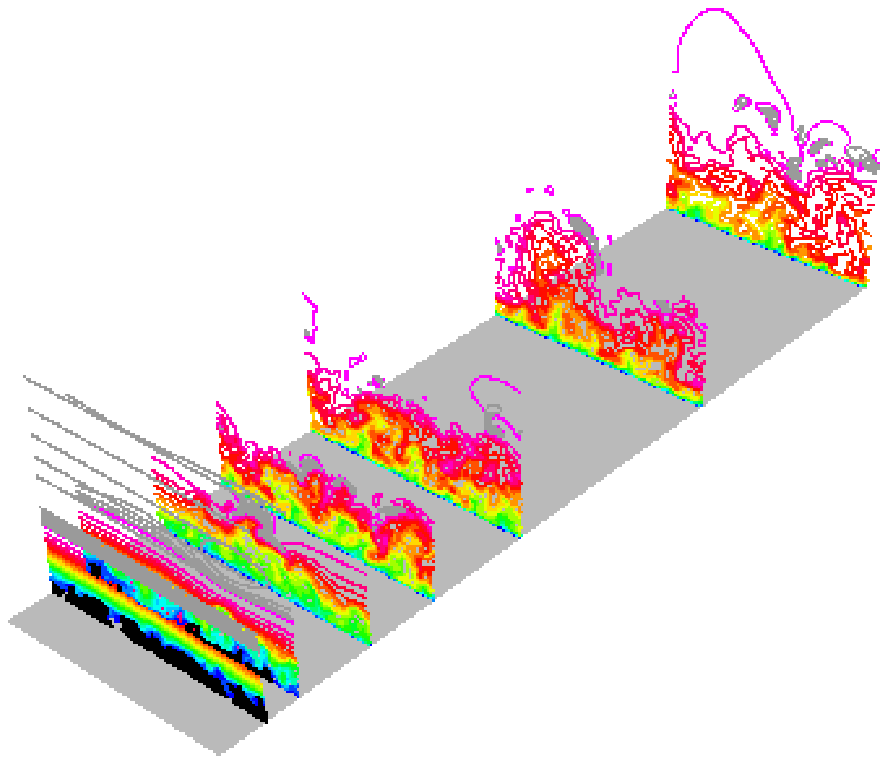}
\vskip -0.3cm \hskip 4cm a) \hskip 7cm b) 
\vskip -0.5cm
\hskip -1.5cm
\includegraphics[width=8.0cm]{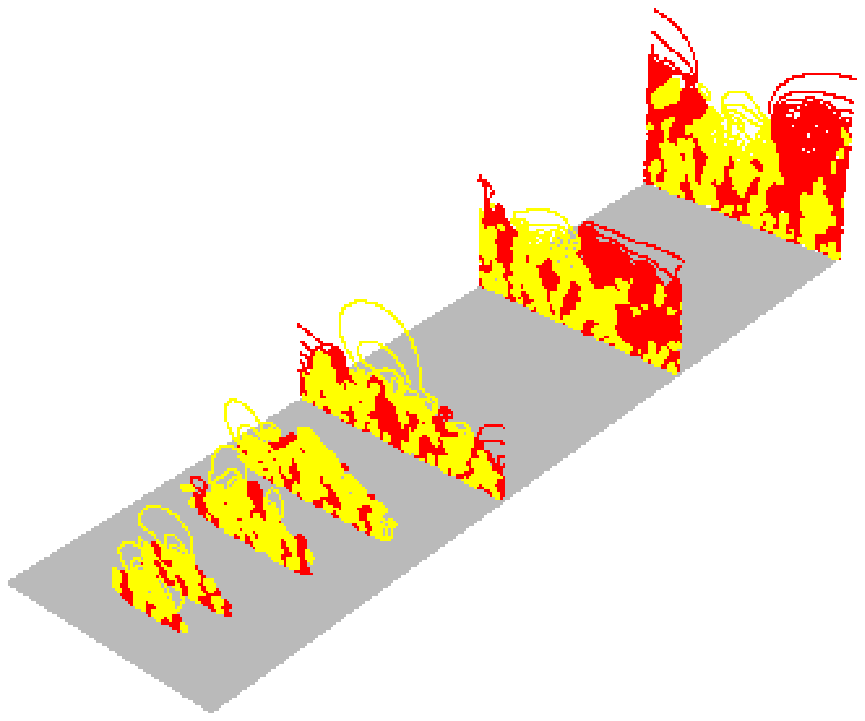}
\hskip -1.2cm
\includegraphics[width=8.0cm]{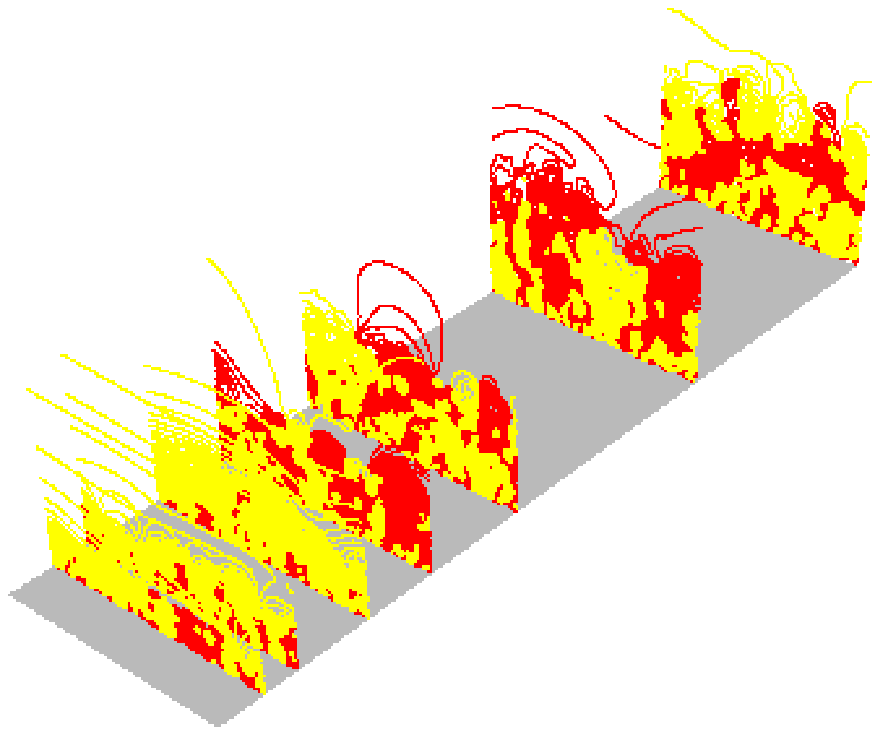}
\vskip -0.3cm \hskip 4cm c) \hskip 7cm d) 
\vskip -0.5cm
\hskip -1.5cm
\includegraphics[width=8.0cm]{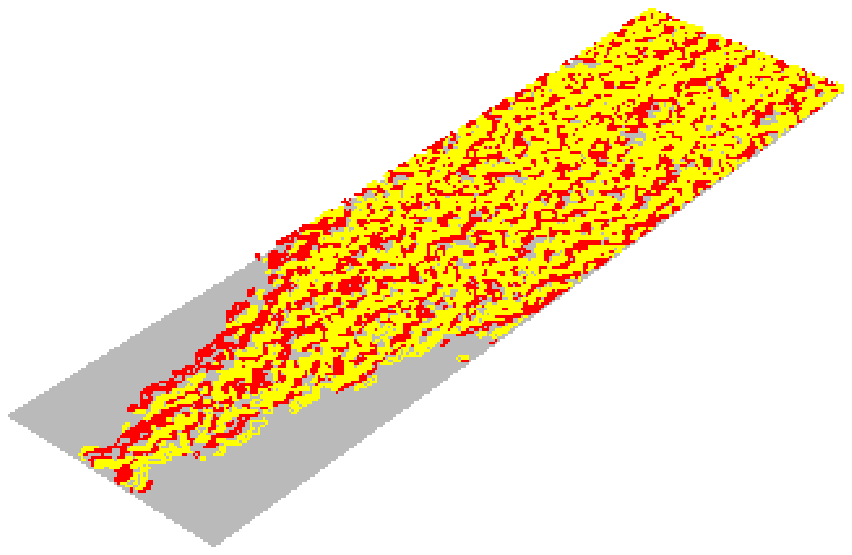}
\hskip -1.2cm
\includegraphics[width=8.0cm]{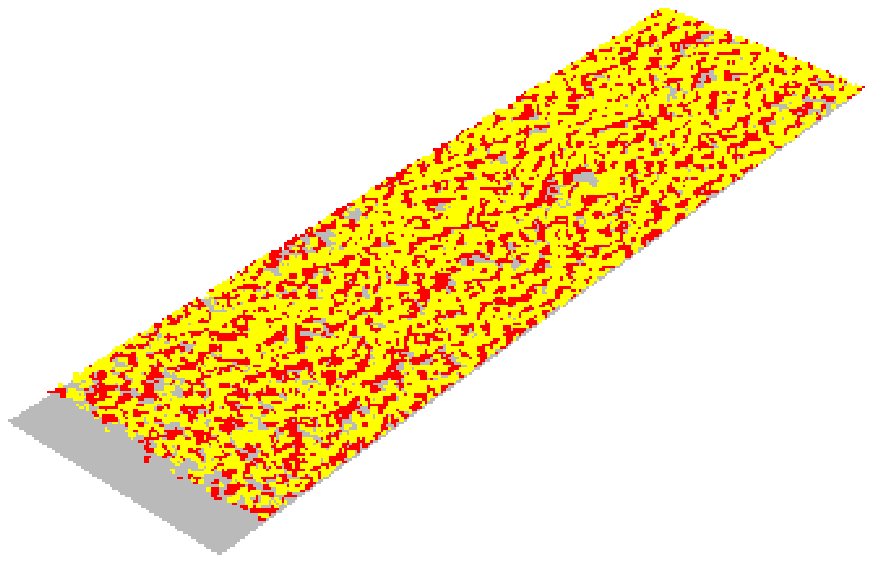}
\vskip -0.3cm \hskip 4cm e) \hskip 7cm f) 
\vskip -0.5cm
\caption{ Contours of a), b) $u_1$ and c), d) $u_2$ at
different $x_2-x_3$ planes  for the cube in the top
figures and for square bars in the bottom at $Re=7500$; $\Delta u_1=0.05$
for positive values $\le 1$ colour from blue to red, 
$\Delta u_1=0.005$ for values $\ge 1$ grey and
black for negative values; $\Delta u_2=0.005$
yellow negative red positive, in e) and f) contours
of $\omega_2$ at $x_2=0.05$ in a $x_1-x_3$ plane
yellow negative and red positive with $\Delta \omega_2=1$.
}
\label{fig10}
\end{figure}

As previously mentioned, the reason of the different trends
to the turbulent regime for two- and three-dimensional obstacles is
linked to the different type of instability. This is qualitatively described 
by the velocity contours at different distances from the obstacle.
Figure \ref{fig10}a shows that the no-slip condition on the side of the cube 
produces two low-speed streaks, better depicted by the $\omega_2$
contours in a horizontal plane at a distance $x_2=0.05$,
in wall units approximately equal to $17$ (figure 9e).  
Further downstream other vortex couples form, which imply 
an array of low and high speed streaks. In the
first two $x_2-x_3$ planes, in figure 9a, 
the undulation at the edge of the boundary
layer is almost negligible, indicating small fluctuations of $u_2$, 
quantified by $u_2$ contours in figure 9c. These disturbances
at the fourth and fifth sections are enough strong to deform
the $u_1$ contours at the edge of the boundary layer in figure 9a.
The $\omega_2$ contours in figure 9e show the end of the
spreading at the fifth section and the  moderate $u_2$ ejections, until there,
are depicted in figure 9c. This figure, further downstream, shows coherent 
ejections going from the near wall region up to the edge of the 
boundary layer. These are the large structures that recently attracted the
interest of a large number of researchers cited  by Pirozzoli \etal \cite{PBO}.
A more quantitative 
analysis of the effects of the large 
scale structures has been described by Pirozzoli \etal \cite{PBO} 
for Couette flows where the turbulent kinetic energy
production, in the central region, facilitates the formation of the large
structures similar to those in the boundary layers.

The transitional region is different for the square bar,
the black region in the contour plots in figure \ref{fig10}b
highlights the large separation bubble which does not reattach
uniformly in agreement with the contours in the second $x_3-x_2$
plane.  The comparison between the contours at the fourth 
plane in figure \ref{fig10}c and figure \ref{fig10}d stresses
large undulations in the outer layer for  the two-dimensional obstacle,
due to the fast growing disturbances
promoted by the inflectional point in the profiles of $u_1$,
located in the black region in figure \ref{fig10}b.
The  contours of $u_2$ in the fourth $x_2-x_3$ plane in figure \ref{fig10}d
confirm the strong interaction between the near wall and the outer regions. 
This observation is corroborated by the wider region where the  profiles of 
$\tup2$, for wall bounded flows  are constant with respect to that
for $\tilde{u}_1^\prime$ and $\tilde{u}_3^\prime$.  
The contours of $\omega_2$ in figure \ref{fig10}f
show  that the near wall streaks close to reattachment line are short, and 
long in the fully turbulent region.

\begin{figure}
\centering
\vskip 0.0cm
\hskip -1.0cm
\psfrag{ylab}{\large $ U^+ $}
\psfrag{xlab}{\large $y^+     $ }
\includegraphics[width=6.5cm]{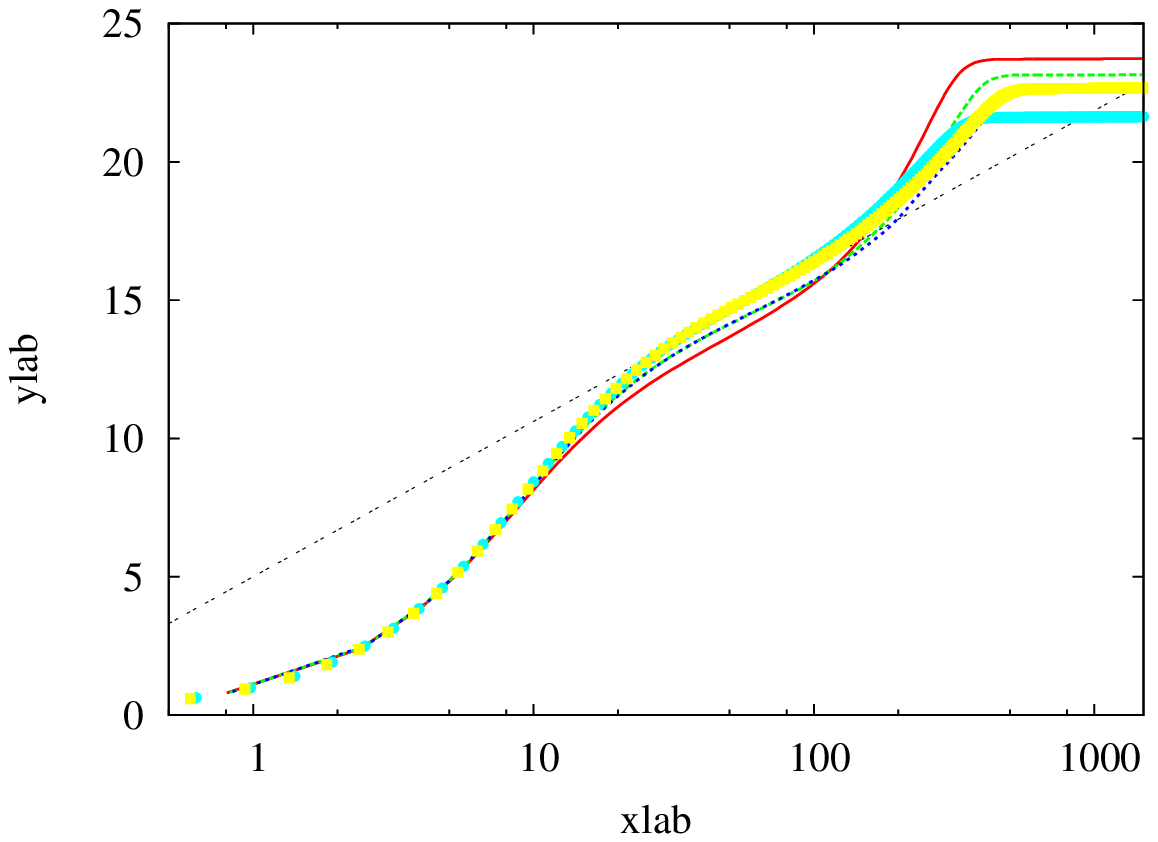}
\hskip 0.0cm
\psfrag{ylab}{\large $  $ }
\psfrag{xlab}{\large $y^+     $ }
\includegraphics[width=6.5cm]{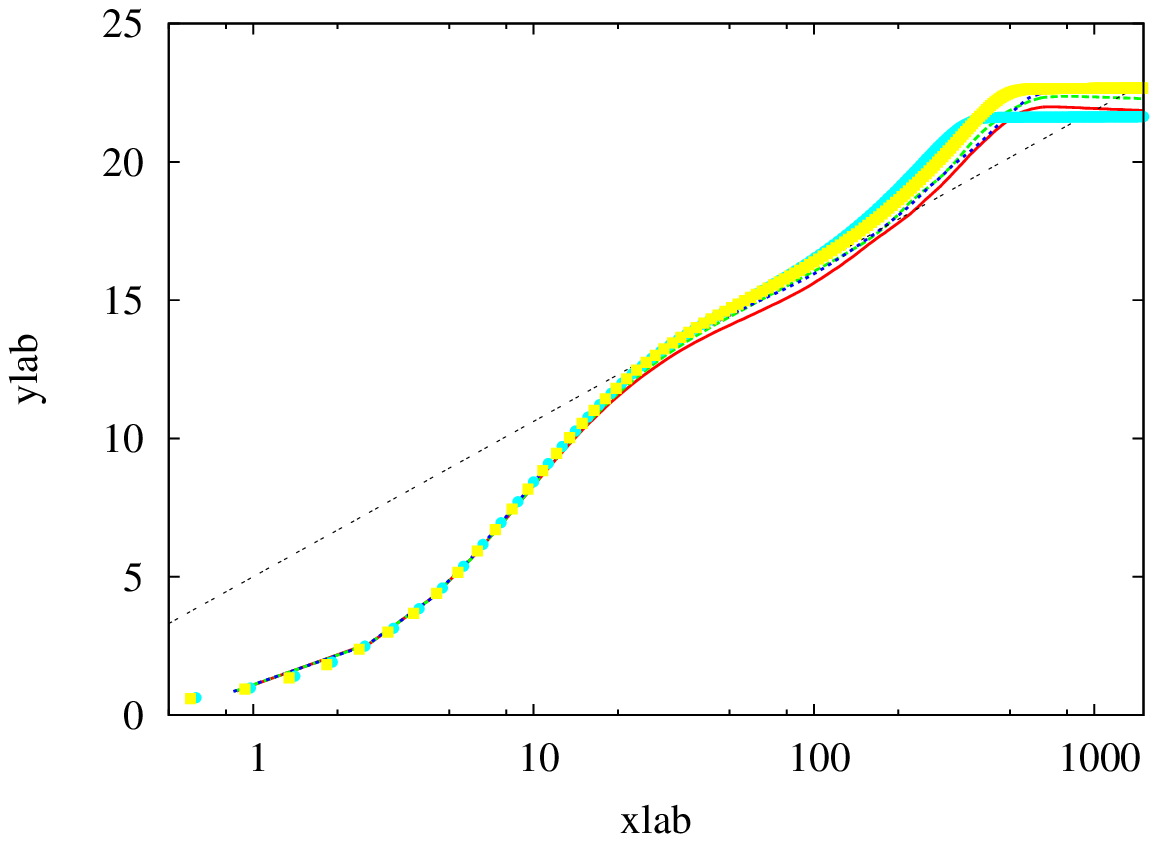}
\vskip -0.3cm \hskip 5cm a) \hskip 5cm b)
\caption{  Mean streamwise velocity profile in wall units:
a) for the cube ($CU$), b) for the square bar ($SQ$)
at three stations (lines) and compared with  the (solid symbols)
Schlatter \etal \cite{SLBJH}
data at $R_\tau = 1000$ and $R_\tau = 1400$.
}
\label{fig11}
\end{figure}

To have a quantitative measure of the distance necessary
to get the fully turbulent regime  the profiles in wall units may 
help.  In boundary layers the correct averaging is in time 
and in the  spanwise  direction.  A further averaging in $x_1$ for a 
dimensionless length equal to $0.1$, accounting for approximately
$50$ profiles, allows
to store less fields.  The profiles 
are calculated at three distances from the obstacle:
equal to $115 k$, $140k$  and $165 k$. In this range 
$1200 > R_\theta > 1400$ therefore the data of Schlatter \etal \cite{SLBJH}
at $R_\theta = 1000$ and $R_\theta = 1400$ are considered.
Figure \ref{fig11}a shows that for the cube at the first station
(red line) the profile of $U^+$ does not have a satisfactory $\log$
law, further downstream the agreement is better, but the profiles
fit the $\log$ law with $B=4.8$ instead of $B=5.0$.  For the 
square bar a much better agreement is observed in Figure \ref{fig11}b where 
only at $x_1=115 k$ the correct $\log$ law is  not fitted. 
The mean velocity profiles approach faster the fully turbulent one, 
however a fully turbulent regime is established when all the statistics 
have the appropriate profile.

\begin{figure}
\centering
\vskip 0.0cm
\hskip -1.0cm
\psfrag{ylab}{\large $ \tilde{u}_1^\prime $}
\psfrag{xlab}{\large $y^+     $ }
\includegraphics[width=6.5cm]{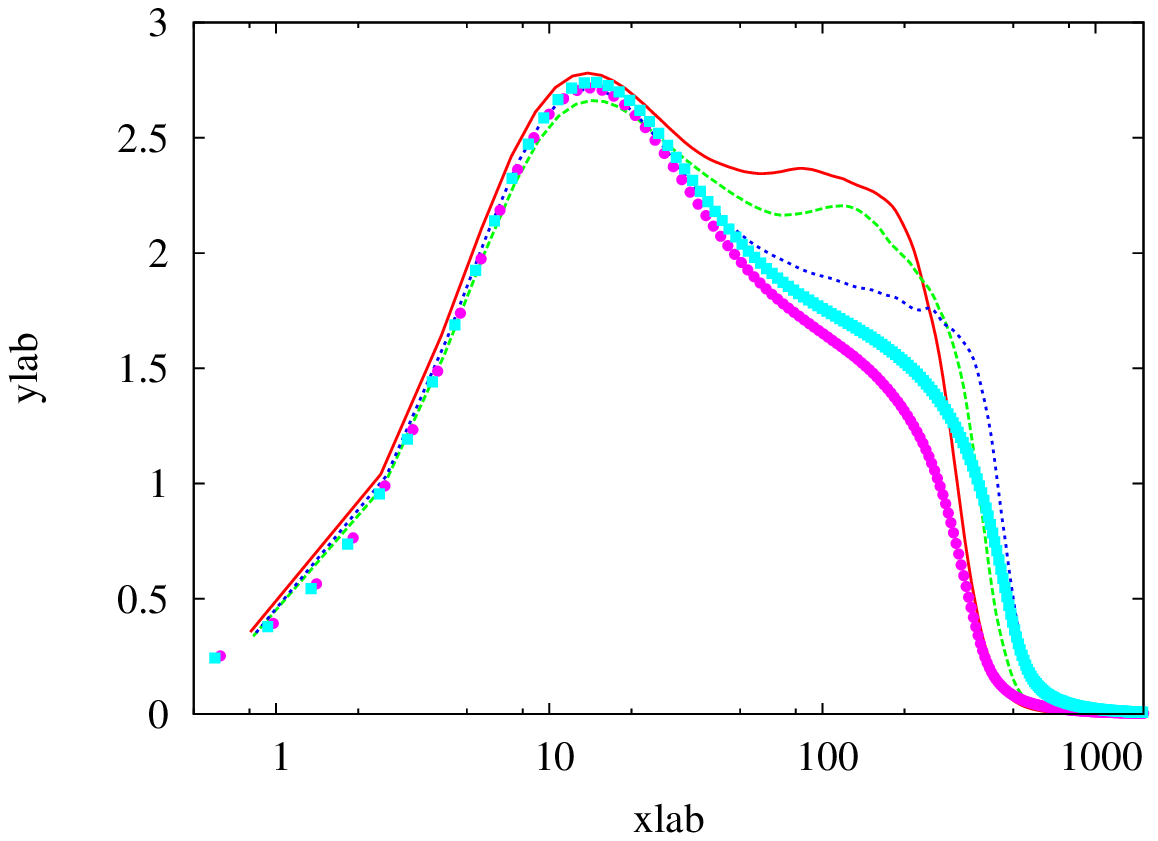}
\hskip 0.0cm
\psfrag{ylab}{\large $  $ }
\psfrag{xlab}{\large $y^+     $ }
\includegraphics[width=6.5cm]{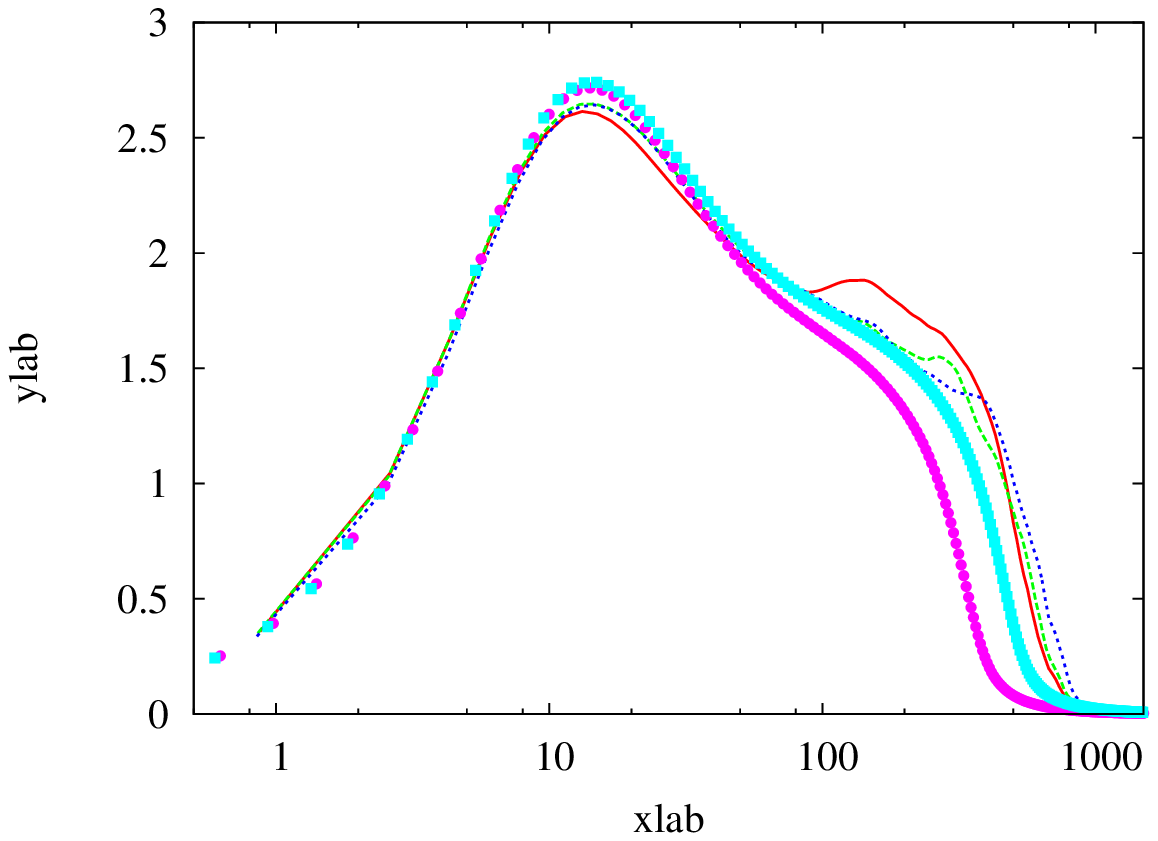}
\vskip -0.5cm \hskip 5cm a) \hskip 5.5cm b)
\vskip 0.0cm
\hskip -1.0cm
\psfrag{ylab}{\large $ \tilde{u}_2^\prime $}
\psfrag{xlab}{\large $y^+     $ }
\includegraphics[width=6.5cm]{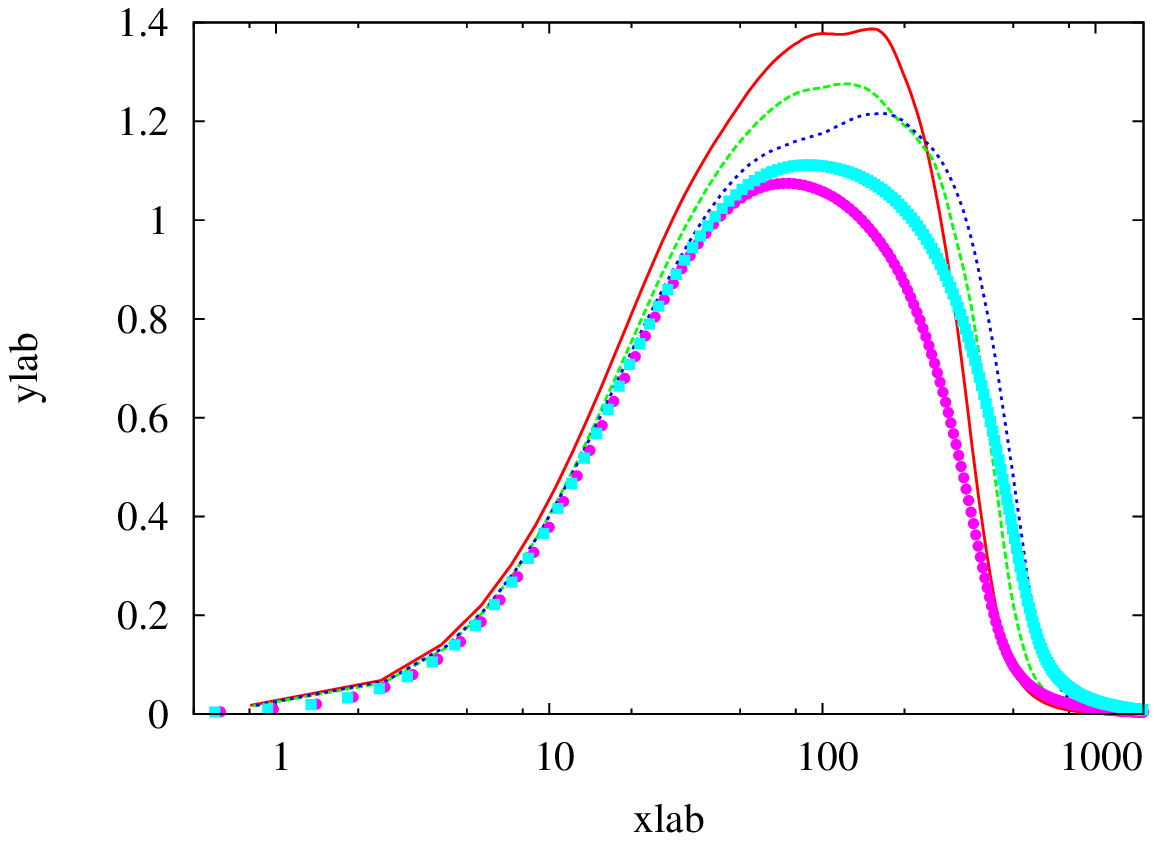}
\hskip 0.0cm
\psfrag{ylab}{\large $  $ }
\psfrag{xlab}{\large $y^+     $ }
\includegraphics[width=6.5cm]{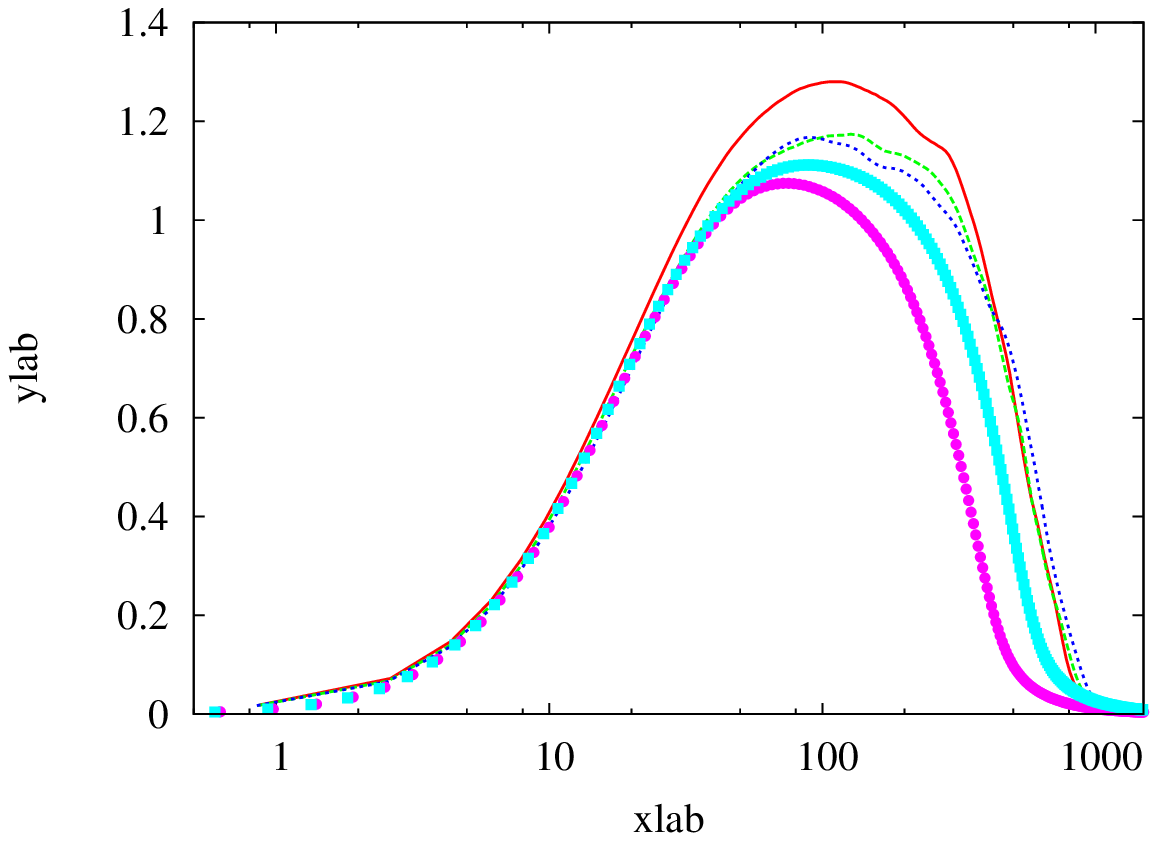}
\vskip -0.5cm \hskip 5cm c) \hskip 5.5cm d)
\vskip 0.0cm
\hskip -1.0cm
\psfrag{ylab}{\large $ \tilde{u}_3^\prime $}
\psfrag{xlab}{\large $y^+     $ }
\includegraphics[width=6.5cm]{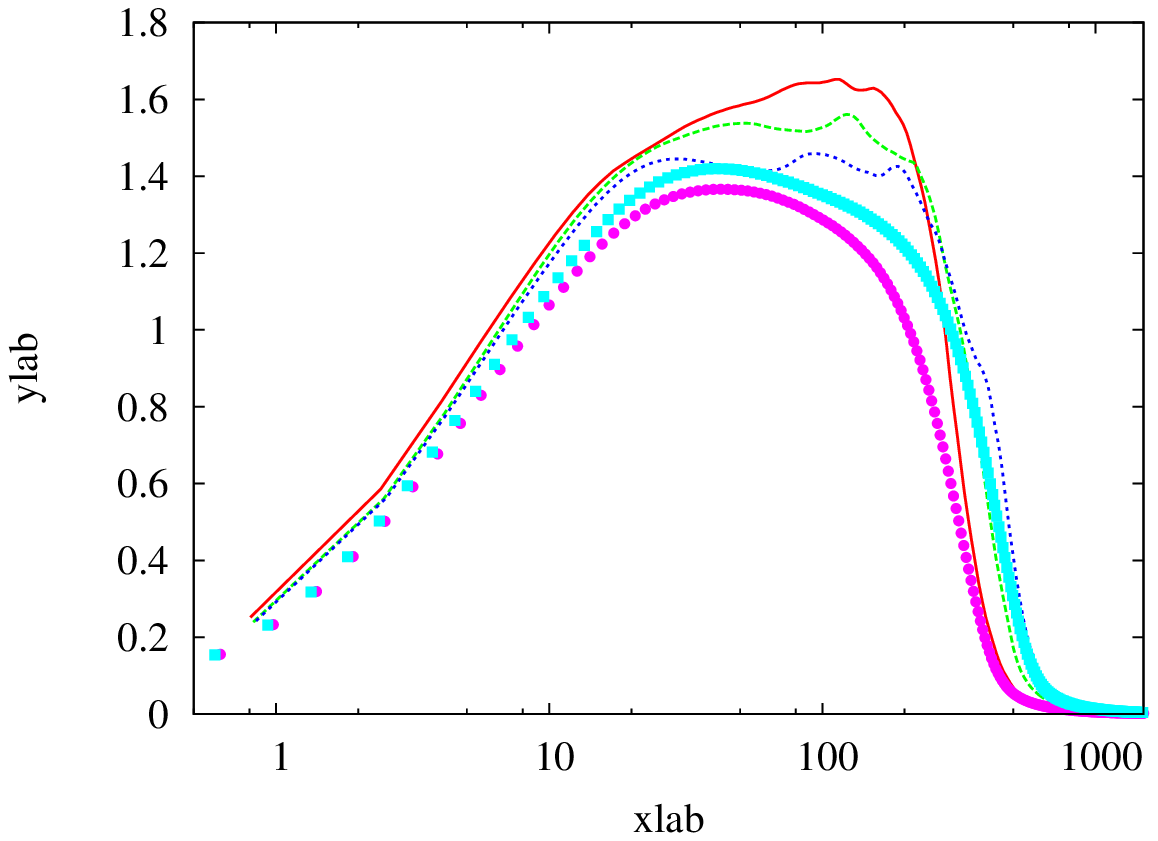}
\hskip 0.0cm
\psfrag{ylab}{\large $  $ }
\psfrag{xlab}{\large $y^+     $ }
\includegraphics[width=6.5cm]{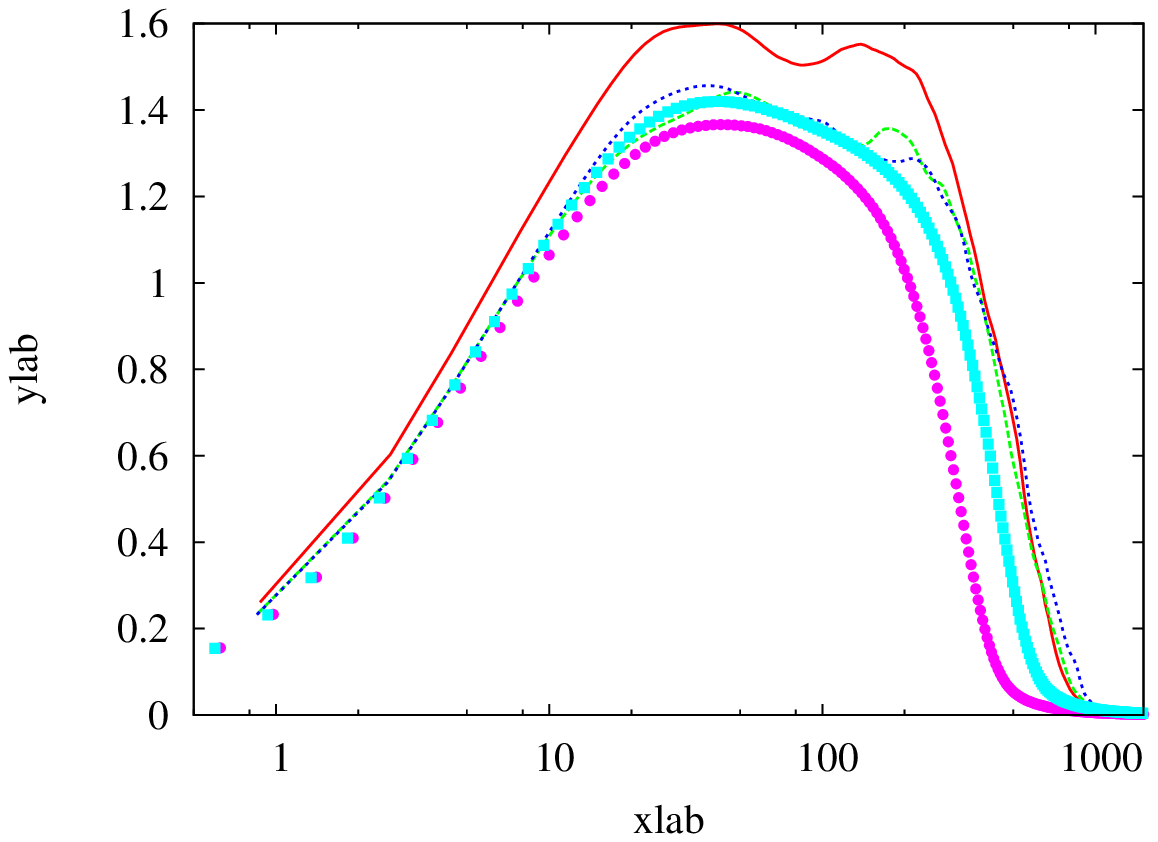}
\vskip -0.5cm \hskip 5cm e) \hskip 5.5cm f)
\caption{  Normal stress   velocity profiles in wall units:
a), c), e) for the cube ($CU$), b), d), f) for the square bar ($SQ$)
at three stations (lines) and compared with  the (solid symbols)
Schlatter \etal \cite{SLBJH}
data at $R_\tau = 1000$ and $R_\tau = 1400$.
}
\label{fig12}
\end{figure}

In figure \ref{fig12} the profiles of the normal stresses at the same 
locations of the $U^+$ in figure 11 reinforce the observation that the fully 
turbulent condition is reached at a shorter distance for the two-           
than for the three-dimensional obstacles. The profiles in   
the near wall region converge  faster implying that the near wall structures 
form closer to the disturbance, and that the large structures in the $\log$ 
and in the outer regions are affected by the inlet conditions. The comparison 
among the three stresses highlights large deviations of $\tilde{u}_2^\prime$
from the correct value, and this fact corroborates our view that, in wall 
bounded flows, this stress accounts for the changes at the wall, as
it was shown by Orlandi \etal \cite{OLTA} and for the disturbances affecting 
the outer region.

\begin{figure}
\centering
\vskip 0.0cm
\hskip -1.0cm
\psfrag{ylab}{\large $ U^+ $}
\psfrag{xlab}{\large $    $ }
\includegraphics[width=6.5cm]{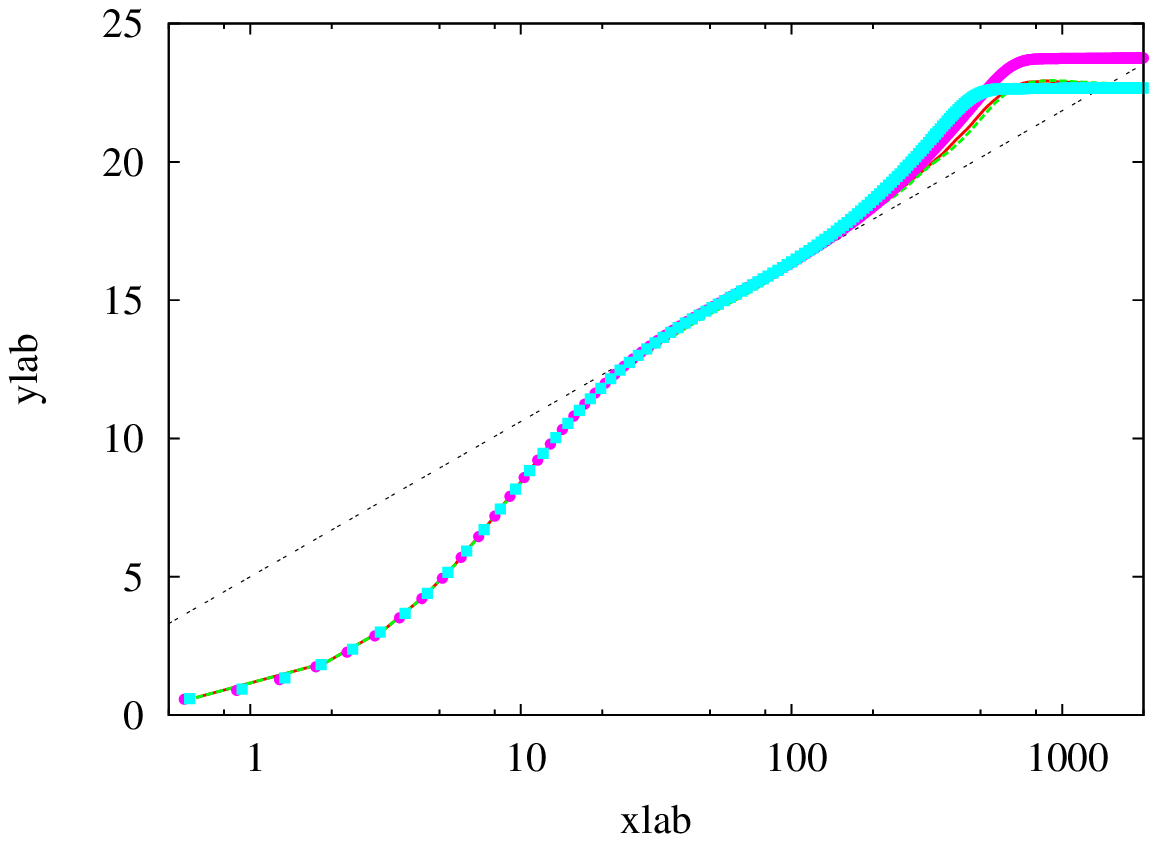}
\hskip 0.0cm
\psfrag{ylab}{\large $ \tilde{u}_1^\prime $ }
\psfrag{xlab}{\large $     $ }
\includegraphics[width=6.5cm]{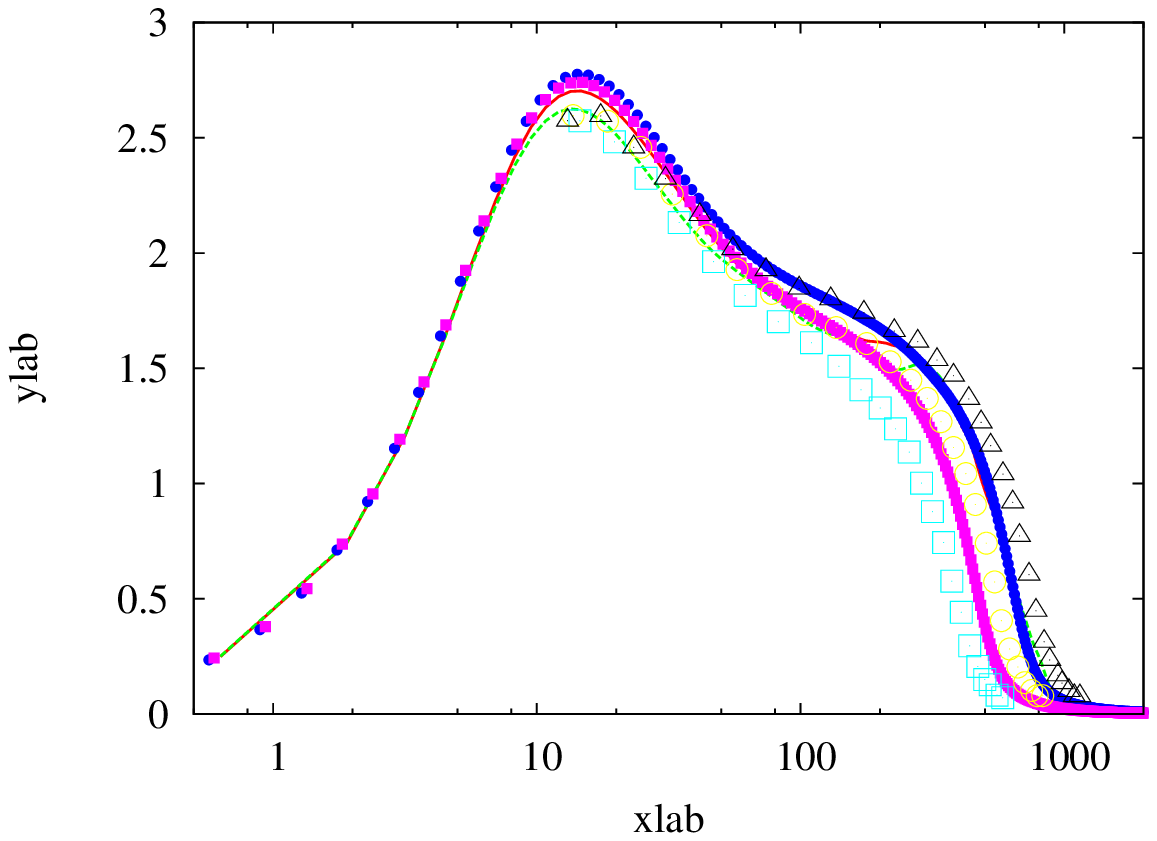}
\vskip -0.5cm \hskip 5cm a) \hskip 5.5cm b)
\vskip 0.0cm
\hskip -1.0cm
\psfrag{ylab}{\large $ \tilde{u}_2^\prime $}
\psfrag{xlab}{\large $y^+     $ }
\includegraphics[width=6.5cm]{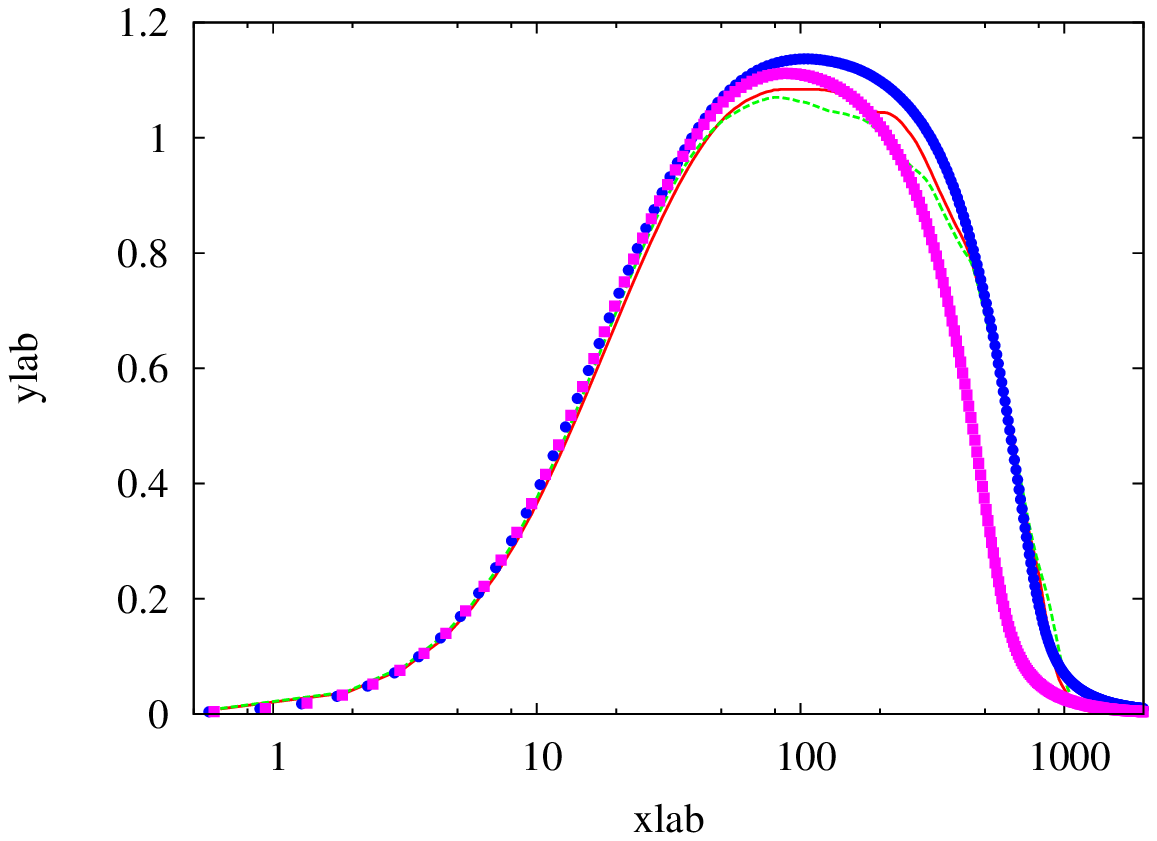}
\hskip 0.0cm
\psfrag{ylab}{\large $ \tilde{u}_3^\prime $ }
\psfrag{xlab}{\large $y^+     $ }
\includegraphics[width=6.5cm]{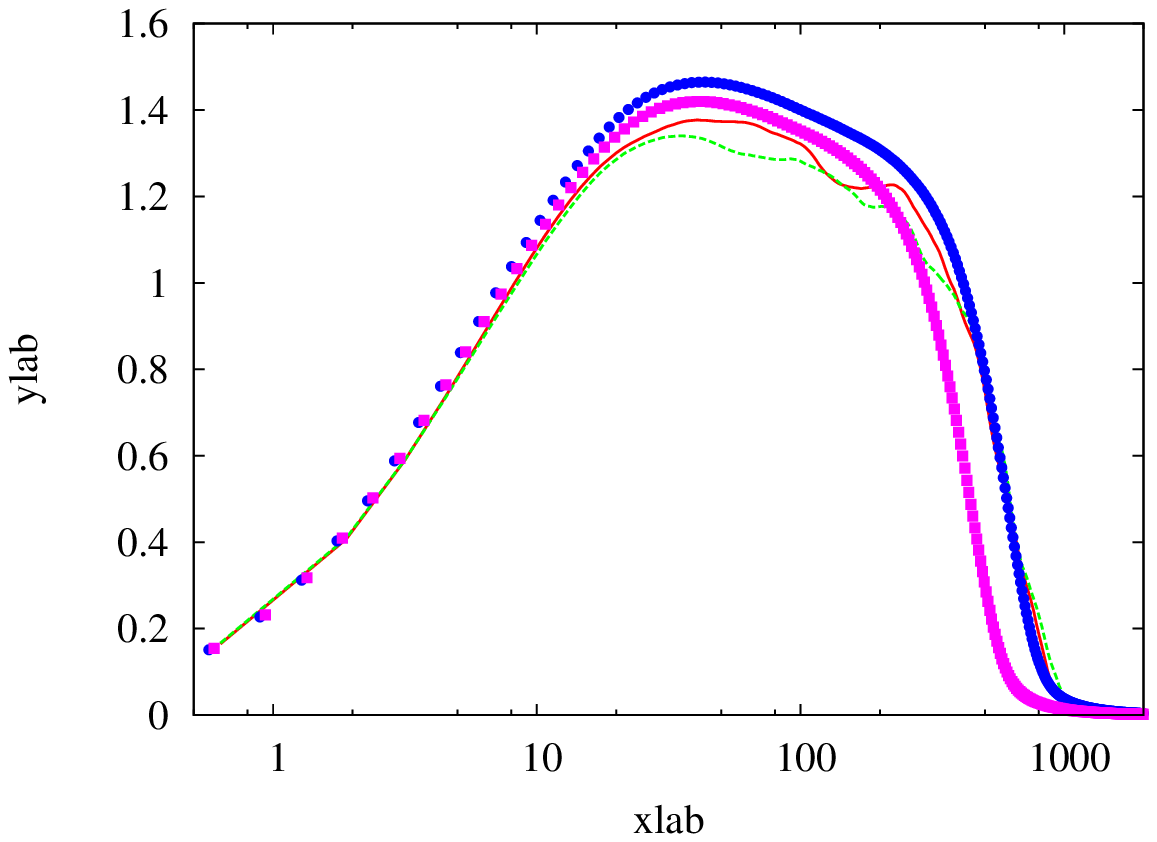}
\vskip -0.5cm \hskip 5cm c) \hskip 5.5cm d)
\caption{ a) Mean streamwise velocity
and b), c), d) normal stress   velocity profiles in wall units 
for the square bar ($SQ$)
at two stations (lines) ($R_\tau \approx 1600$) compared with  the (solid symbols)
Schlatter \etal \cite{SLBJH} 
data at $R_\tau = 1000$ and $R_\tau = 2000$.
and the Erm \& Joubert \cite{EJ} (open symbols) experiments
at $R_\tau = 1000, 1500$ and $R_\tau = 2000$.
}
\label{fig13}
\end{figure}

Figure 12 shows large differences between the 
Schlatter \etal \cite{SLBJH} results and the present ones. 
The reason is related to the 
insufficient $L_1$ to get the fully turbulent regime, to the
smaller resolution than  that in Schlatter \etal \cite{SLBJH}
and to the different type of disturbances.
Therefore, a more refined simulation, with the parameters previously 
reported, was performed for the square bar, having observed to
be the most efficient to promote the transition.  This simulation is 
a first attempt to reproduce the most complete and documented experiments 
in turbulent boundary layers by Erm \& Joubert \cite{EJ}. The reproduction of 
the small obstacles used in the experiments requires an enormous amount of 
computational time , therefore in the discussion it should be recalled that 
the height of the tripping device ($k/\delta=0.25$)
is greater than that in the experiment ($k/\delta=0.1$). 
This difference together with the limitations in $L_1$ 
affect the wake region.  Far from the obstacles 
$R_\theta\approx 1600$ is a value intermediate between the 
$R_\theta=1000$ and $R_\theta=2000$ by Schlatter \etal \cite{SLBJH}.
The very good agreement for $U^+$ is appreciated in figure \ref{fig13}a. 
Near the wall the stresses (figure \ref{fig13}b,d)
are slightly smaller than those by Schlatter \etal \cite{SLBJH}; 
the  reason may be related to the different  resolution.
The small oscillations in the outer regions,
in large part, are due to the large scales produced by the obstacle.
In a longer domain these oscillations disappear, but to show it
a very large CPU is required.  These oscillations, due to the trip, 
are not encountered in simulations based on a recycling procedure, where 
these long-lived structures, in the outer region, are not continuously 
introduced.  
At $R_\theta \approx 700$, Erm \& Joubert \cite{EJ2} claimed that 
differences with the Spalart \cite{Spa} DNS were related to the tripping.  
In the experiments the height  was  $0.1\delta$, therefore, 
with an obstacle of height $0.25\delta$  it is reasonable 
to find in the outer layer differences between the present and the 
results obtained with the recycling procedure.

\begin{figure}
\centering
\vskip 0.0cm
\hskip -1.5cm
\psfrag{ylab}{\large $ E_{1}^* $}
\psfrag{xlab}{\large $     $ }
\includegraphics[width=7.0cm]{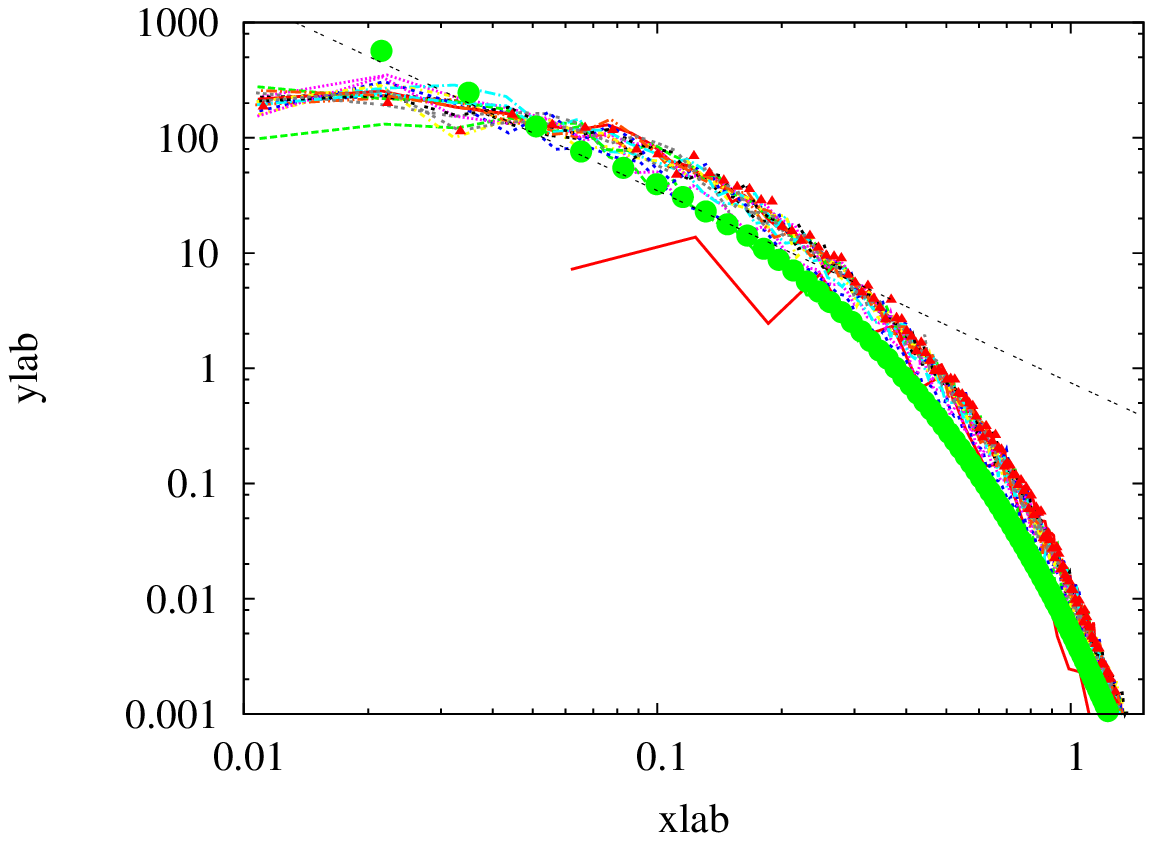}
\hskip -1.0cm
\psfrag{ylab}{\large $     $ }
\psfrag{xlab}{\large $     $ }
\includegraphics[width=7.0cm]{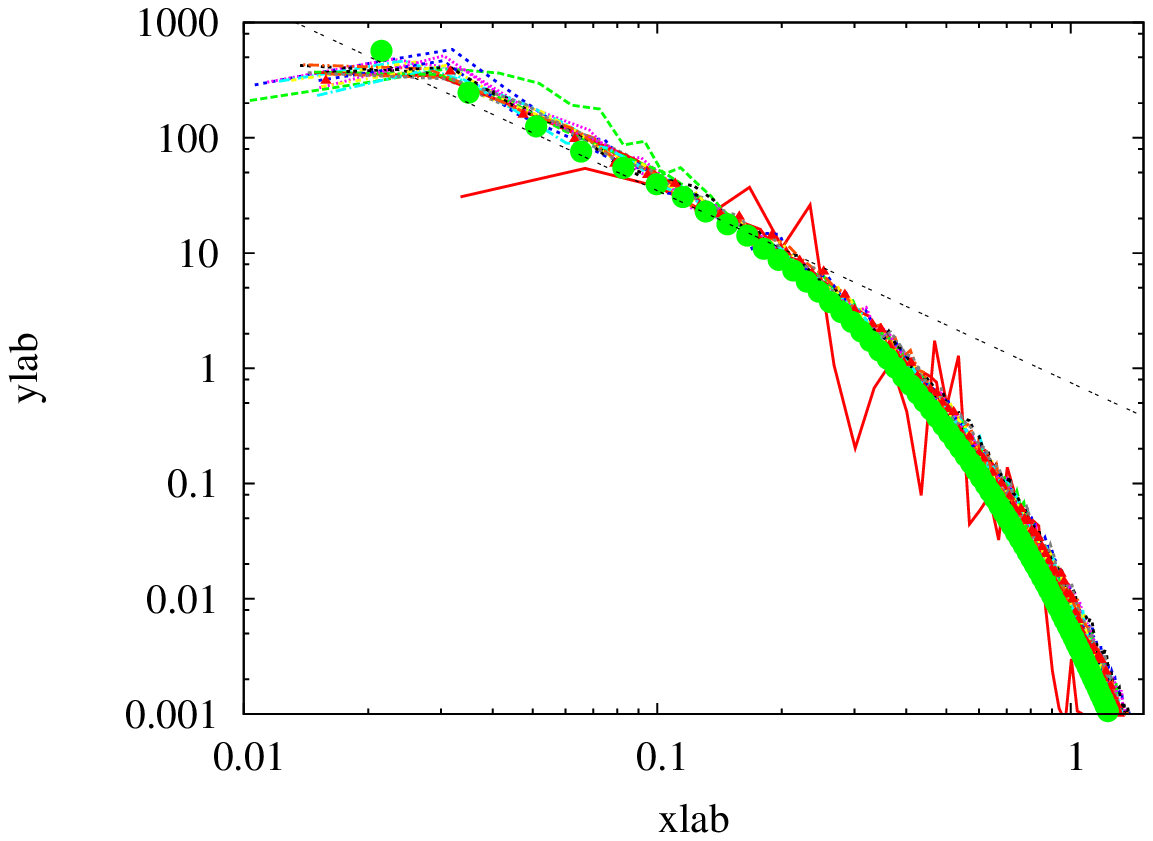}
\vskip -0.3cm \hskip 0cm a) \hskip 6cm b) 
\vskip 0.0cm
\hskip -1.5cm
\psfrag{ylab}{\large $ E_{2}^* $}
\psfrag{xlab}{\large $     $ }
\includegraphics[width=7.0cm]{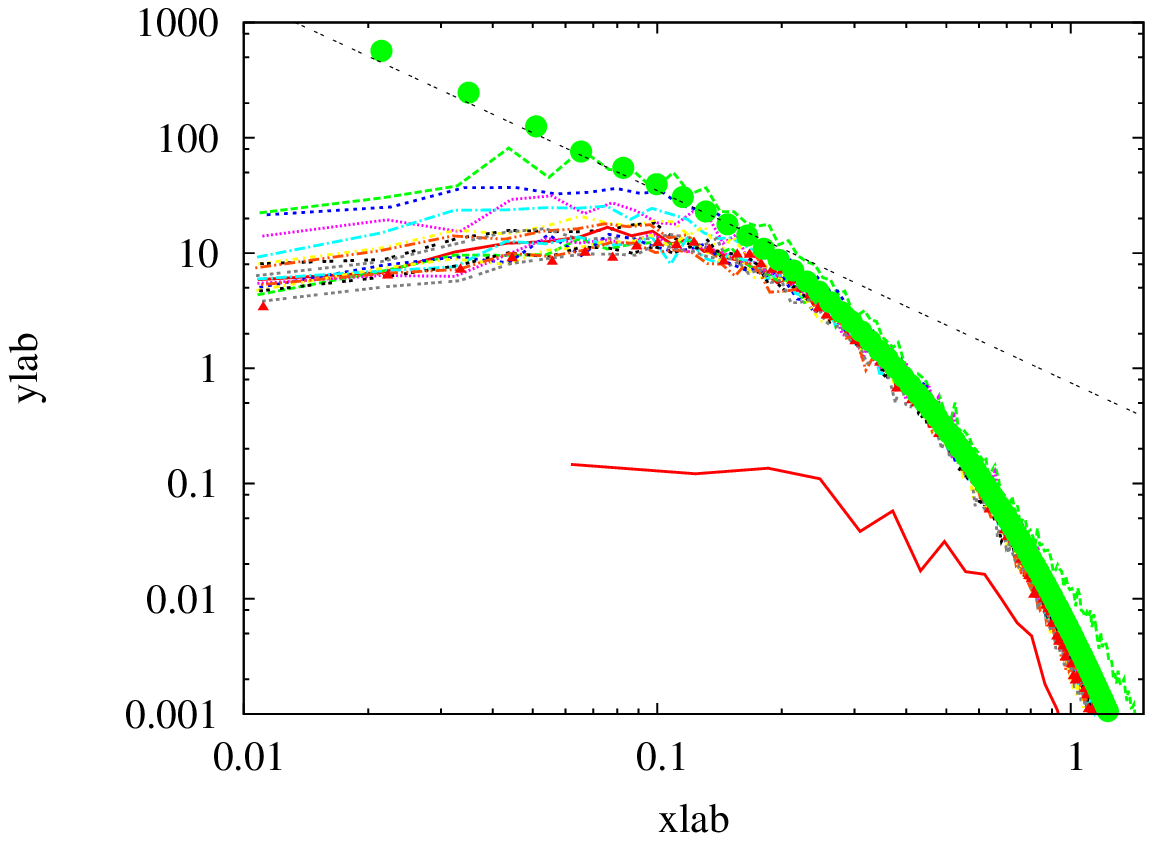}
\hskip -1.0cm
\psfrag{ylab}{\large $     $ }
\psfrag{xlab}{\large $     $ }
\includegraphics[width=7.0cm]{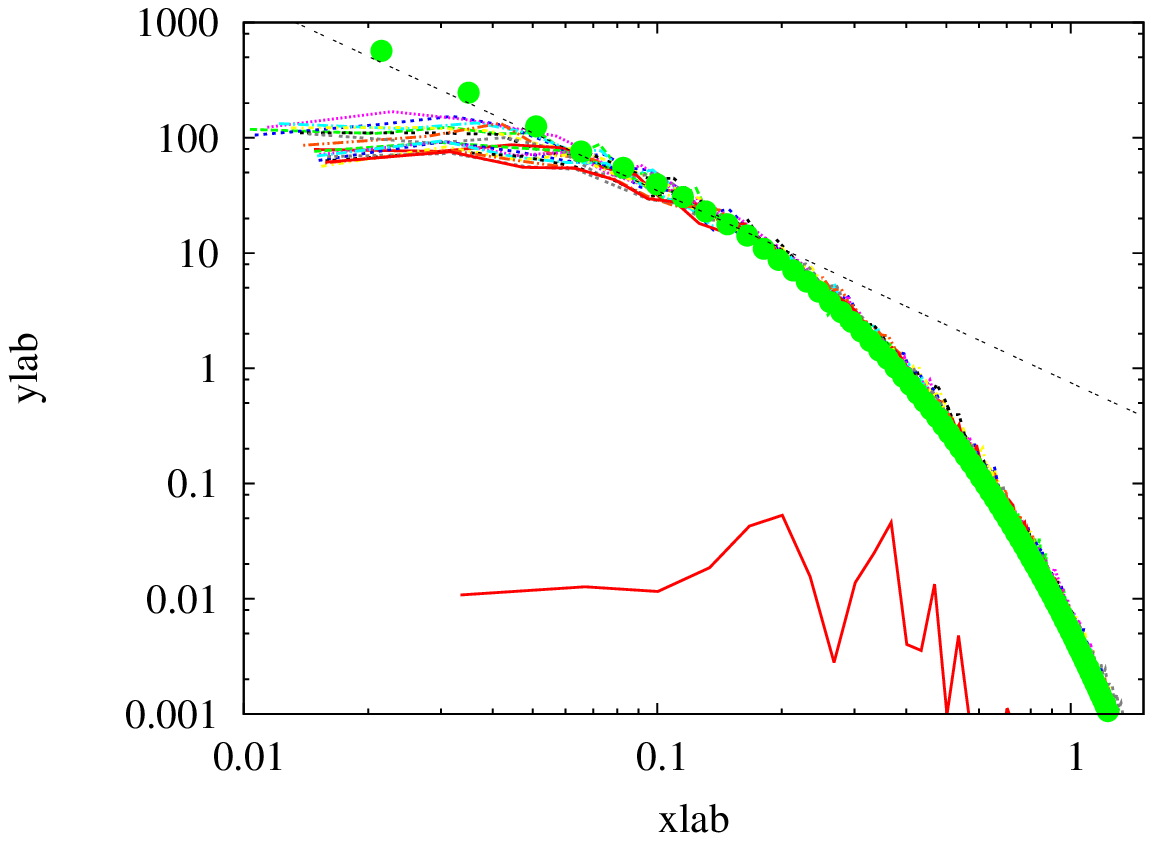}
\vskip -0.3cm \hskip 0cm c) \hskip 6cm d) 
\vskip 0.0cm
\hskip -1.5cm
\psfrag{ylab}{\large $ E_{3}^* $}
\psfrag{xlab}{\large $k_3^*     $ }
\includegraphics[width=7.0cm]{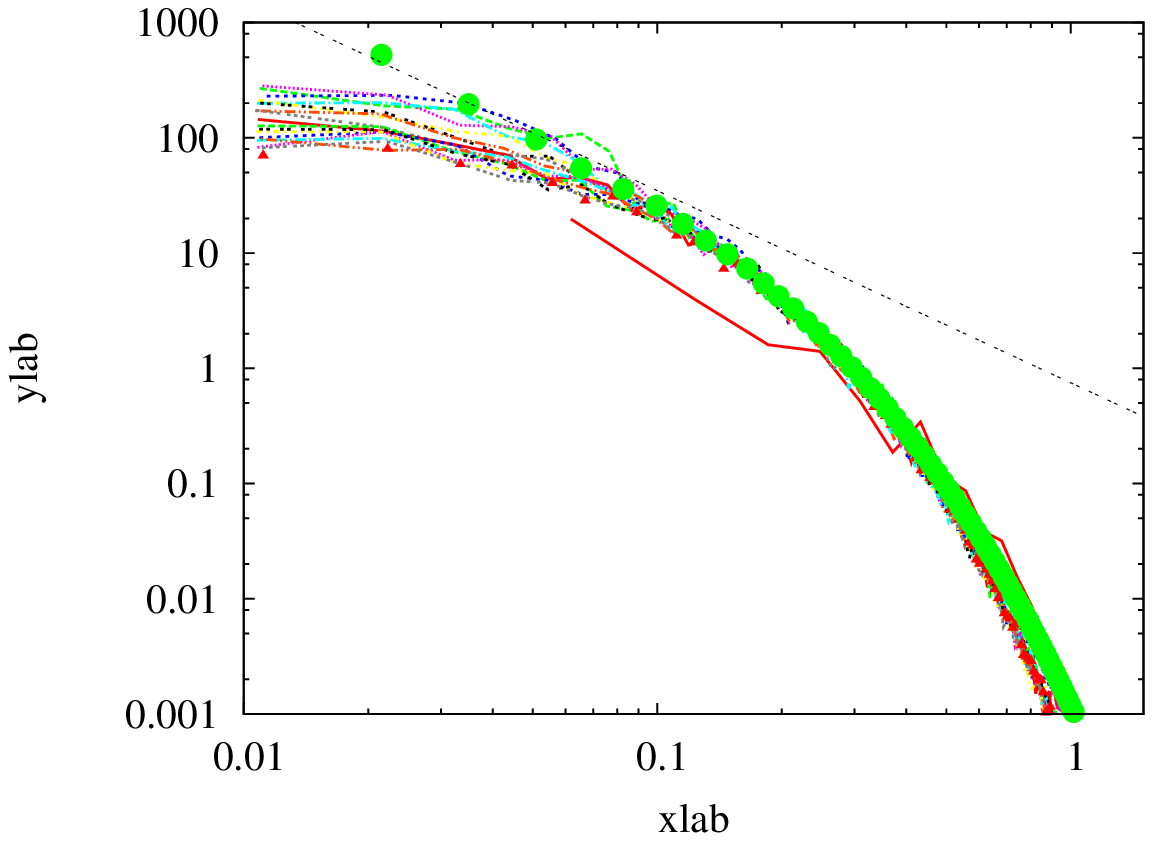}
\hskip -1.0cm
\psfrag{ylab}{\large $     $ }
\psfrag{xlab}{\large $k_3^*     $ }
\includegraphics[width=7.0cm]{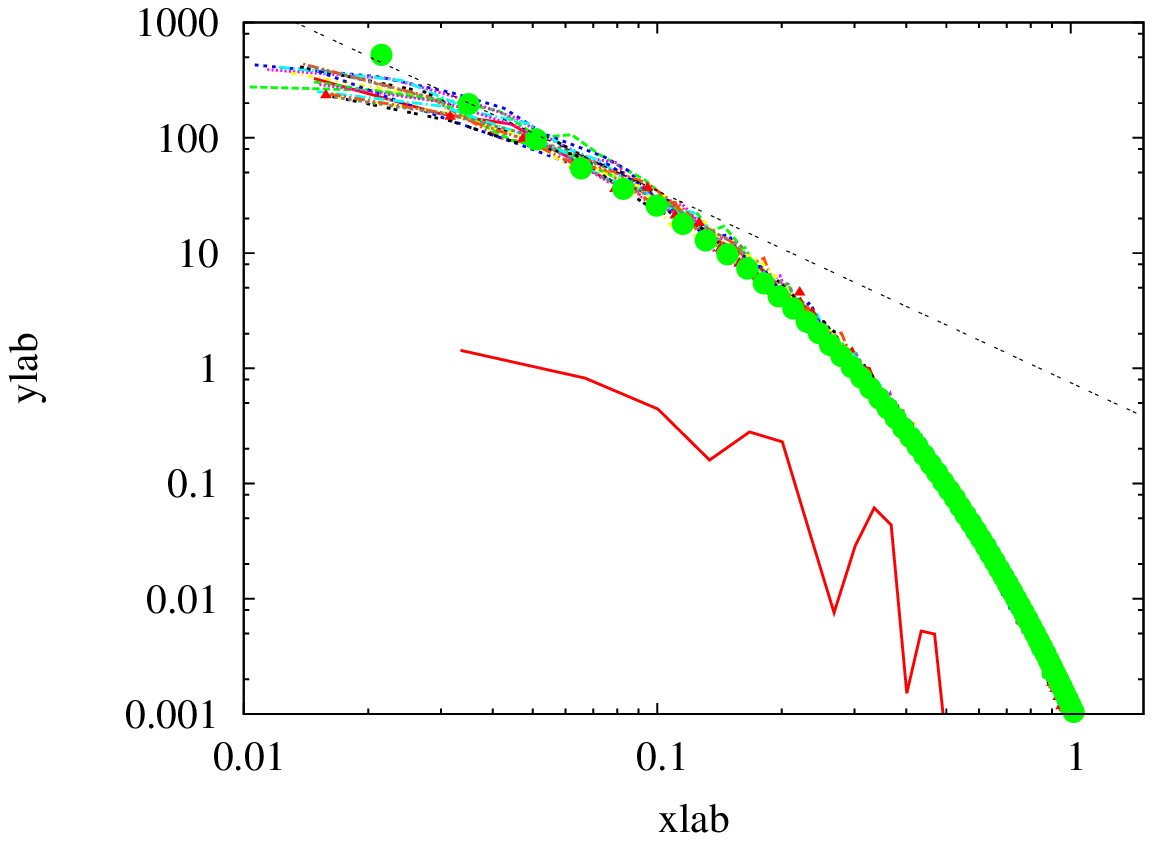}
\vskip -0.3cm \hskip 0cm e) \hskip 6cm f) 
\vskip 0.0cm
\caption{  Spectra in kolmogorov units a), c), e) at $y^+=25$
, b), d), f) at $y^+=110$ in the spanwise direction (lines), 
compared with isotropic turbulence
at $R_\lambda=100$ (Jimenez \etal \cite{JWSR}) open symbols,
a) , b) ,c), d) transverse
e), f) longitudinal.
}
\label{fig14}
\end{figure}

To demonstrate that the turbulent fields are fully resolved 
it is worth looking at the spectra in Kolmogorov units, if a
collapse in the exponential range with those of isotropic turbulence occurs
it means that a true DNS is performed. 
The agreement in the whole range implies
the reproduction of the turbulent energy cascade process. 
The velocity spectra in the spanwise direction are compared with the 
longitudinal and transverse spectra by Jimenez \etal \cite{JWSR}
at $R_\lambda=98.7$.  The spectra at $y^+=25$ at different
locations, separated by a distance equal to $3.2\delta$ starting
from $x_1=6.4$, are presented in figure \ref{fig14}.  At the first location,
inside the separation bubble, the spectra are completely different than the 
others. In the transitional and in the fully turbulent regions
a good collapse of the spectra at high $k_3$ is obtained, corroborating
the previous statement that the small scale adjust in a short
distance. In agreement with figure 13 the energy containing scales for $u_1$ 
are those adjusting faster. Instead those for $u_2$ require a longer distance 
to reach the streamwise independence in the fully turbulent regime (indicated
by small triangles in figure 13). The spectra for $u_1$ in the $\log$ region
at $y^+=110$ (the figures on the right of figure \ref{fig14}) 
have a reasonable wide inertial
range. In addition, a better collapse of the spectra of $u_2$ is achieved
(figure \ref{fig14}d), and the tendency towards isotropic turbulence
is emphasised in figure \ref{fig14}d and figure \ref{fig14}e.
From the computational side the last two figures 
are a convincing proof that these simulations are true DNS,    
having resolved scales of size close to the Kolmogorov scales.

\section{Conclusions}
\label{conclusion}

One common way to promote the transition of laminar boundary 
layers is through tripping devices inserted at a certain distance
from the leading edge. A large number of
experiments were devoted to find a critical Reynolds number,
but a systematic study of the influence of the shape of the
obstacle was never attempted, even if it was requested by
Klebanoff \etal \cite{KCT}. This lack of data motivated
the present study and from the results achieved it was clear
that a critical Reynolds number based on the velocity at the
top of the obstacle ($U_k$) and on its height $k$ could not be
the satisfactory quantity marking the transition. In fact it is reasonable
to expect that by varying the shape of the obstacle, $U_k$ could not
drastically change. This has been confirmed by the present simulations,
but in the mean time it  was observed that for certain shapes
there was transition and for others not.  In agreement with the 
previous results in rough channels (Orlandi \cite{O11}) it has been
observed that the transition occurs when $\tup2|^+_k $  is greater than  
a threshold value. In the channel 
the periodic conditions lead to an easy determination of the threshold value.
In boundary layer the determination is more critical for the streamwise 
evolution of $\tup2|_k $. However the numerical simulations 
demonstrate that, in analogy to the case of a normal wall jet where the 
transition is linked to the exit velocity, also for solid obstacles
if the amplitude of $\tup2|_k $ is greater than a threshold limit the transition
occurs.

The DNS furnishes the vorticity field, and flow visualizations
of the three vorticity components lead to conclude that 
$\omega_2$ is the most important,  modifying the $\omega_3$
sheet generated at the top of the obstacle. The head of the 
horseshoe vortex generated by the obstacle does not contribute 
to the transition. It has been, also understood that for
isolated elements the transition is enahnced by geometries
which create curved vorticity sheets. To reach a fully turbulent conditions
the wake generated by isolated obstacles expand laterally by
the generation of opposite sign vorticity at
the wall. The entire process can be summarised as the wake instability 
mentioned by Klebanoff \etal \cite{KCT}, which differs from the inflectional
instability occurring in presence of two-dimensional obstacles. The present
DNS demonstrated that the latter instability is
more efficient to produce  a fully turbulent regime, and  
therefore simulations with square bars have been performed 
at high $Re$ and the results 
have been compared with previous numerical simulations, and
laboratory experiments.

The most refined simulation, with a two-dimensional trip, produced mean 
velocity and turbulent stresses profiles in a good agreement with
those achieved in DNS with recycling procedures or with ad hoc
inlet disturbances.  The results suggest that the numerical method
here used is an efficient tool to create a
data base with all the ingredients necessary to understand the differences
among wall bounded flows in pipe, channels and boundary
layers. In the simulations "hot wires" measuring the three velocity,
and vorticity components, together with the pressure have been inserted.
The temporal date can be used to understand
the differences on the spectra obtained by the Taylor hypothesis 
and those evaluated in the spanwise directions. 
The latter spectra were presented as a proof of the
reproduction of the physics of turbulent flows characterised
by an universal exponential range.  
The temporal signals may help to understand the 
quality of a hot wire, having observed that the smallest Kolmogorov scales
have been captured in the numerical simulations. 

The present numerics is not appropriate to reproduce tripping
devices of smaller height, requiring large computational
resources. To achieve this goal, with limited resources, 
modifications of the numerics with non-uniform
grids in all three directions are necessary.
In this way the increased resolution
near the obstacle allows to create the vorticity sheets driving
the transition. For incompressible flows this is a difficult task,
but it is feasible with non-uniform grids in $x_1$ and $x_2$.
Our aim is to introduce these modifications 
in order to reproduce the experiments
by Erm \& Joubert \cite{EJ}  with a two-dimensional trip 
device with $k/\delta=0.1$. For three-dimensional obstacles
a good clustering can be obtained by using the compressible
code at low Mach number used by Bernardini \etal \cite{BPO}.
In these circumstances the small time steps require
a much large  computational time than that used for the simulations
here discussed.

\section{Acknowledgments}

The support of a MIUR 60 \% grant is acknowledged.
The computer time was given by CASPUR and by CINECA.

\end{document}